\documentclass[12pt]{article}
\pdfoutput=1

\usepackage[bulletsep]{collref}
\usepackage{amssymb,graphicx}
\usepackage[intlimits]{amsmath}
\usepackage{bbm}
\usepackage[small]{subfigure}

\usepackage{MnSymbol}

\makeatletter \@addtoreset{equation}{section} \makeatother

\makeatletter
\let\old@startsection=\@startsection
\let\oldl@section=\l@section
\renewcommand{\@startsection}[6]{\old@startsection{#1}{#2}{#3}{#4}{#5}{#6\mathversion{bold}}}
\renewcommand{\l@section}[2]{\oldl@section{\mathversion{bold}#1}{#2}}
\makeatother

\makeatletter
\let\old@makecaption=\@makecaption
\def\@makecaption{\small\old@makecaption}
\makeatother

\renewcommand{\leq}{\leqslant}

\newcommand{\sfrac}[2]{{\textstyle\frac{#1}{#2}}}
\newcommand{\half}{\sfrac{1}{2}}
\newcommand{\ihalf}{\sfrac{i}{2}}

\newcommand{\neel}{{\rm N\acute{e}el}}
\newcommand{\mps}{{\rm MPS}}
\newcommand{\ur}{\uparrow}
\newcommand{\dr}{\downarrow}

\begin{document}


\setcounter{page}{1}
\renewcommand{\thefootnote}{\arabic{footnote}}
\setcounter{footnote}{0}

\begin{titlepage}

\begin{flushright}\footnotesize
\texttt{NORDITA-2015-72} \\
\texttt{UUITP-12/15}
\vspace{0.6cm}
\end{flushright}

\centerline{\large \bf One-point Functions in Defect CFT and Integrability}
\vskip 2 cm

\centerline{  {\bf Marius de Leeuw$\,^{1}$},  {\bf Charlotte Kristjansen$\,^{1}$} and {\bf Konstantin Zarembo$\,^{2}$}  }

\vskip 1.1cm

\begin{center}
\sl $^1$ The Niels Bohr Institute, University of Copenhagen \\
\sl  Blegdamsvej 17, DK-2100 Copenhagen \O , Denmark
\vskip 0.3cm
\sl $^2$ NORDITA, KTH Royal Institute of Technology and Stockholm University \\
Roslagstullsbacken 23, SE-106 91 Stockholm, Sweden  \\
Department of Physics and Astronomy, Uppsala University\\
SE-751 08 Uppsala, Sweden
\end{center}
\vskip 0.7cm

\centerline{\small\tt deleeuwm@nbi.ku.dk, kristjan@nbi.ku.dk, zarembo@nordita.org}

\vskip 1.7cm \centerline{\bf Abstract} \vskip 0.2cm \noindent

\noindent
We calculate planar  tree level one-point functions \mbox{of\,}
non-protec\-ted
operators in the defect conformal field 
theory dual to the D3-D5 brane system with $k$ units of the world volume flux.  
Working in the operator basis of Bethe eigenstates of the  Heisenberg $XXX_{1/2}$ spin chain we express the one-point functions
as  overlaps of these  eigenstates with a matrix product state.  For $k=2$ we obtain a closed expression
of determinant form for  any number of excitations, and in the 
case of half-filling we find a relation with the N\'eel state.
In addition, we present a number of results for the limiting case $k\rightarrow  \infty$.

\end{titlepage}


\section{Introduction}
The simplest probes of external heavy objects  in a conformal field theory, such as Wilson or 't~Hooft lines, surface operators or interfaces, are one-point functions of local operators in the presence of the defect. By conformal symmetry,
\begin{equation}\label{<O>}
 \left\langle \mathcal{O}(x)\right\rangle=\frac{C}{z^\Delta }\,,
\end{equation}
where $z$ is the distance  from $x$ to the defect and $\Delta $ is the scaling  dimension of the operator $\mathcal{O}$. The constant $C$ in principle depends on the normalization of the operator at hand, but if the two-point function of $\mathcal{O}$ is unit-normalized, $C$ is defined unambiguously.

Here we focus on a domain wall in $\mathcal{N}=4$ Super-Yang-Mills (SYM) theory which  separates vacua with $SU(N)$ and $SU(N-k)$ gauge groups  \cite{DeWolfe:2001pq}. This defect originates from the D3-D5 brane intersection and is dual to a probe D5 brane in $AdS_5\times S^5$ with $k$ units of electric flux on its world-volume \cite{Karch:2000gx}. One-point functions of chiral operators in this \cite{Nagasaki:2012re} and in the closely related D3-D7 defect CFT \cite{Kristjansen:2012tn}, when continued to strong coupling perfectly agree with the predictions of the AdS/CFT duality. 

We would like to make a connection with integrability and will thus consider expectation values of non-protected operators. It has proven useful in this context to study operators of large bare dimension, which correspond to long quantum spin chains. Conformal  operators of this type, due to operator mixing, are linear combinations of a large number of field monomials. Efficient calculation of the classical expectation values for such operators becomes a non-trivial problem, which can only be solved by employing the full machinery of the  Bethe ansatz. The one-point correlators are probably the simplest objects sensitive to the structure of the Bethe wavefunctions, and are thus ideally suited to probe integrability beyond the spectral data.

\section{Domain wall and spin chains}\label{sec:Defect}

The D3-D5 intersection defect in $\mathcal{N}=4$ SYM  has the following semiclassical description at weak coupling. On the one side of the domain wall, the gauge symmetry is broken from $SU(N)$ to $SU(N-k)$ by an infinite scalar vev.  On the other side the scalar fields relax to zero according to their classical equations of  motion:
\begin{equation}\label{2ndordereqm}
 \frac{d^2\Phi^{\rm cl} _i}{dz^2}=\left[\Phi^{\rm cl} _j,\left[\Phi^{\rm cl} _j,\Phi^{\rm cl} _i\right]\right].
\end{equation}
For a supersymmetric defect, the solution also satisfies the first-order Nahm equations \cite{Nahm:1979yw}:
\begin{equation}
 \frac{d\Phi ^{\rm cl}_i}{dz}=\frac{i}{2}\,\varepsilon _{ijk}\left[\Phi ^{\rm cl}_j,
 \Phi ^{\rm cl}_k\right],
\end{equation}
which automatically imply (\ref{2ndordereqm}).
The solution describing the D3-D5 intersection is \cite{Diaconescu:1996rk,Giveon:1998sr,Constable:1999ac}:
\begin{equation}\label{Phiclass}
 \Phi _i^{\rm cl}=\frac{1}{z}\,\begin{pmatrix}
  \left(t_i\right)_{k\times k} & 0_{k\times (N-k)} \\ 
  0_{(N-k)\times k} & 0_{(N-k)\times (N-k)} \\ 
 \end{pmatrix},~i=1,2,3,\qquad \Phi ^{\rm cl}_i=0,~ i=4,5,6,
\end{equation}
where the three $k\times k$ matrices $t_i$  satisfy
\begin{equation}\label{eq:su2relations}
 \left[t_i,t_j\right]=i\varepsilon _{ijk}t_k,
\end{equation}
and consequently
realize the unitary $k$-dimensional representation of $\mathfrak{su}(2)$.

The one-point functions, to the first approximation, are obtained by simply replacing quantum fields in the operator with their classical expectation values \cite{Nagasaki:2012re,Kristjansen:2012tn}. To get a non-zero answer the operators must be built from scalar fields, and we will consider the most general such operators that do not contain derivatives:
\begin{equation}\label{operator}
 \mathcal{O}=\Psi ^{i_1\ldots i_L}\mathop{\mathrm{tr}}\Phi _{i_1}\ldots \Phi _{i_L}.
\end{equation}
The  $SO(6)$ tensor $\Psi $ is cyclically symmetric because of the trace condition. 

These operators form a closed sector at one loop, and their mixing is described by an integrable $SO(6)$ spin-chain Hamiltonian, wherein the tensor $\Psi $ plays the role of the wave function in the spin-chain Hilbert space. 
 The anomalous part of the dilatation generator (the mixing matrix) at one loop  contains only nearest-neighbor interactions \cite{Minahan:2002ve}:
\begin{equation}\label{dilop}
 \Gamma =\frac{\lambda }{16\pi ^2}\sum_{l=1}^{L}H_{l,l+1},\qquad 
 H_{lm}=2-2P_{lm}+K_{lm},
\end{equation}
where $\lambda=g^2N $ is the 't~Hooft coupling of the SYM theory, and $P_{lm}$ and $K_{lm}$ are permutation and trace operators acting on sites $l$ and $m$ of the spin chain:
\begin{equation}
 P_{ij}^{ks}=\delta ^k_j\delta ^s_i,\qquad K_{ij}^{ks}=\delta_{ij}\delta ^{ks}.
\end{equation}
This result is not modified by the presence of the defect \cite{DeWolfe:2004zt}. Notice, however,
that the latter reference deals with a probe brane set-up without fluxes (corresponding to $k=0$) but the ultraviolet 
divergencies of the theory should be the same when the classical fields are turned on.

The Hamiltonian (\ref{dilop}) is a member of an infinite hierarchy of commuting charges responsible for the integrability of the model. The third charge\footnote{According to the standard convention the first charge is the momentum along the spin chain and the second charge is the Hamiltonian itself.} of the hierarchy acts on three neighboring spins:
\begin{equation}\label{Q3}
 Q_3=\sum_{l=1}^{L}Q_{l},\qquad Q_{l}=[H_{l-1,l},H_{l,l+1}].
\end{equation}
Unlike the Hamiltonian, the third charge is parity-odd, and changes sign under the inversion of the spin chain orientation\footnote{This symmetry is equivalent to charge conjugation in SYM.}.  The spectrum of the spin chain therefore contains parity pairs with degenerate energy and opposite values of $Q_3$, as well as unpaired states with vanishing $Q_3$.

The defect CFT contains also operators localized on the domain wall. These operators are described by an integrable open spin chain \cite{DeWolfe:2004zt}  and are dual to open strings with ends attached to the D5 brane. By considering one-point functions of the bulk operators we are, in a sense, dealing with the same string diagram but viewed as an absorption of a closed string by the D5 brane. In string theory the two descriptions should be related by $t-s$ channel duality, and it would be interesting to understand how they are related at weak coupling.

By substituting (\ref{Phiclass}) into (\ref{operator}) we  find that the one-point function is proportional to
\begin{equation}\label{eq:overlap}
 \Psi ^{i_1\ldots i_L}\mathop{\mathrm{tr}}t_{i_1}\ldots t_{i_L}\equiv \left\langle 
 {\rm MPS}\,\right.\!\!\left|\vphantom{\Psi ^{\rm cl}}\Psi \right\rangle,
\end{equation}
the inner product of the spin-chain state $\Psi $ that characterizes the operator and the state with the wave function 
\begin{equation}\label{MPSwavefunction}
 {\rm MPS}_{i_1\ldots i_L}=\mathop{\mathrm{tr}}t_{i_1}\ldots t_{i_L}.
\end{equation}
MPS here stands for the 'Matrix Product State', the term that will be explained below. 
The defect thus maps to a particular {\it state} in the spin-chain Hilbert space. We may interpret this state as a weak-coupling counterpart of the boundary state that describes the D5 brane in closed string theory.
Recovering the normalization factor that makes the bulk two-point function of $\mathcal{O}$ unit-normalized, we get for the  structure constant:
\begin{equation}\label{genericCso6}
 C=
 \left(\frac{8\pi ^2}{\lambda }\right)^{\frac{L}{2}}L^{-\frac{1}{2}}\,
 \frac{\left\langle {\rm MPS}\,\right.\!\!\left|\vphantom{{\rm MPS}}\Psi \right\rangle}{\left\langle \Psi \right.\!\!\left|\Psi  \right\rangle^{\frac{1}{2}}}.
\end{equation}

What can be said about the state associated with the defect? It is not an eigenstate of the spin-chain Hamiltonian. We do not get anything nice when apply (\ref{dilop}) to (\ref{MPSwavefunction}). However, the third charge of the integrable hierarchy acts in a simple way and actually annihilates the defect state:
\begin{equation}\label{Q3|>=0}
Q_3 \left|{\rm MPS} \right\rangle=0.
\end{equation}
The proof is given in appendix~\ref{Q3-appendix}. This property leads to a selection rule for the one-point functions, since
the overlap with MPS vanishes for all states that carry $Q_3\neq 0$.

To further simplify the problem we
 consider the $SU(2)$ subsector composed of operators which are built from two complex scalars 
\begin{eqnarray}
 Z=\Phi _1+i\Phi _4&\longleftrightarrow&\left|\uparrow\right\rangle,
\nonumber \\
W=\Phi _2+i\Phi _5&\longleftrightarrow&\left|\downarrow\right\rangle.
\end{eqnarray}
The $SU(2)$ sector is closed to all loop orders, and at the leading order is described by the Heisenberg spin chain. 

When restricted to the $SU(2)$ sector, the spin-chain state associated with the defect becomes
\begin{equation}\label{MPS}
 \left\langle {\rm MPS}\,\right|=\mathop{\mathrm{tr}}\nolimits_a
 \prod_{l=1}^{L}\left(\left\langle \uparrow_l\right|\otimes t_1
 +\left\langle \downarrow_l\right|\otimes t_2\right).
\end{equation}
The index $a$ is introduced here to distinguish the ``auxiliary" space of color indices of $t_i$ from the quantum space spanned by $\left|\uparrow\right\rangle$, $\left|\downarrow\right\rangle$ on each site of the spin chain.
The defect state \eqref{MPS} can be obtained by applying an operator, which we can call the defect operator, to the ferromagnetic ground state of the spin chain:
\begin{equation}
 \left\langle {\rm MPS}\,\right|=\left\langle \uparrow\ldots \uparrow\right|K.
\end{equation}
The defect operator is not uniquely defined, because there are many operators that annihilate the ground state. We can choose it in the form
\begin{equation}\label{MPOgen}
 K=\mathop{\mathrm{tr}}\nolimits_a\prod_{l=1}^{L}
 \left\{\left[s\mathbbm{1}+(1-s)\sigma _l^3\right]\otimes t_1
 +\sigma ^+_l\otimes t_2+\sigma ^-_l\otimes t\right\},
\end{equation}
where $\sigma ^i_l$ are the Pauli matrices acting on the $l$-th site of the spin chain, $s$ is an arbitrary complex number, and $t$ can be any $k\times k$ matrix.
For instance, taking $s=0$ and $t=t_2$, we find:
\begin{equation}\label{1MPO}
 K=\mathop{\mathrm{tr}}\nolimits_a\prod_{l=1}^{L}
 \left(\sigma _l^3\otimes t_1+\sigma _l^1\otimes t_2\right),
\end{equation}
which takes particularly simple form for $k=2$, with $t_1=\sigma ^3/2$ and $t_2=\sigma ^1/2$:
\begin{equation}\label{2MPO}
 K^{\left(k=2\right)}=2^{-L}\mathop{\mathrm{tr}}\nolimits_a\prod_{l=1}^{L}
 \left(\sigma _l^3\otimes\sigma _a^3+\sigma _l^1\otimes\sigma _a^1\right).
\end{equation}
States of the form (\ref{MPS}) are known as the {\it Matrix Product States}, and were extensively studied in the condensed-matter literature \cite{klumper1991equivalence,klumper1993matrix,accardi1981topics,fannes1989exact,fannes1992finitely,ostlund1995thermodynamic,rommer1997class,Alcaraz:2003ya,Alcaraz:2006bt,verstraete2008matrix,Katsura:2009vx}, in particular to model quantum entanglement in one-dimensional systems. The operators (\ref{MPOgen}), (\ref{1MPO}) and (\ref{2MPO}) are usually called the {\it Matrix Product Operators}.

In analogy to the algebraic Bethe ansatz (ABA)  \cite{Faddeev:1996iy} the  construction of the MPS uses the auxiliary space which threads through all sites of the spin chain. Interestingly, here the auxiliary space has a direct physical meaning of the color $SU(N)$ representation in the underlying gauge theory.

The conformal operators in the $SU(2)$ sector are labelled by zero-momen\-tum eigenstates of the Heisenberg Hamiltonian. In the ABA framework, the eigenfunctions are constructed by applying creation operators $B(u)$ to the ferromagnetic vacuum of the spin chain:
\begin{equation}\label{Bethebasic}
 \left|\left\{u_j\right\}\right\rangle=B(u_1)\ldots B(u_M)\left|0\right\rangle.
\end{equation}
Each $B$-operator flips one spin, and for the state to be an eigenstate of the Heisenberg Hamiltonian the rapidities $\{u_i\}$ must fulfil the set of Bethe equations \cite{Faddeev:1996iy}. Our goal is to calculate the structure constant (\ref{genericCso6}) for an arbitrary Bethe state of the form (\ref{Bethebasic}).

The trace cyclicity of the SYM operators imposes the zero-momentum constraint on the Bethe eigenstates. A simple way to fulfil this condition is to consider states in which rapidities come in pairs (the momentum is an odd function of $u$):
\begin{equation}\label{unpairedst}
 \left|\mathbf{u}\right\rangle=\left|u_1\ldots u_{\frac{M}{2}}\right\rangle\equiv \left|\left\{u_j,-u_j\right\}\right\rangle.
\end{equation}
Of course this way to impose the zero-momentum constraint is too restrictive and there are zero-momentum Bethe states in which rapidities are not balanced pairwise. These states form degenerate parity pairs related by reflection of all rapidities. Such paired states, however, carry a non-zero $Q_3$ and  have zero overlap with the defect state as a consequence of (\ref{Q3|>=0}). We can thus concentrate on the fully balanced, unpaired states of the form (\ref{unpairedst}). 

Our goal is to calculate
\begin{equation}
 C_{\mathbf{u}}=
 \left(\frac{8\pi ^2}{\lambda }\right)^{\frac{L}{2}}L^{-\frac{1}{2}}\,
 \frac{\left\langle {\rm MPS}\,|\mathbf{u}\right\rangle}
 {\left\langle  \mathbf{u}\right.\!\!\left|\mathbf{u}\right\rangle^{\frac{1}{2}}}\,.\label{overlap}
\end{equation}
 There is a considerable literature on overlaps of Bethe states in integrable systems (see \cite{inverse,Slavnov2007} for reviews), which in many cases admit compact determinant representation. The most famous examples are the Gaudin norm of an ABA state \cite{Gaudin:1976sv,Korepin:1982gg}, which is a part of the expression we need to evaluate, and the overlap of the on-shell and off-shell Bethe states \cite{Slavnov:1989}.  Overlaps of Bethe states with MPS have not been studied so far, to the best of our knowledge.
From known results the one that comes closest to our setup is the overlap of an arbitrary Bethe state with the N\'eel state, which was calculated in \cite{Pozsgay:2009} and transformed into a convenient determinant form in \cite{Brockmann:2014a,Brockmann:2014b}.

Bethe-state overlaps are playing an important r\^ole in the gauge/string integrability.
The three-point functions in the $\mathcal{N}=4$ SYM at weak coupling can be expressed as generalized overlaps of Bethe states \cite{Escobedo:2010xs,Escobedo:2011xw,Gromov:2011jh,Gromov:2012vu,Gromov:2012uv,Foda:2013nua,Kazama:2013rya,Jiang:2014mja,Caetano:2014gwa,Kazama:2014sxa,Jiang:2014cya,Basso:2015zoa,Kazama:2015iua} and can be rendered into a compact determinant form \cite{Foda:2011rr,Foda:2012wf,Foda:2012yg,Foda:2012wn,Kostov:2012wv}, which is particularly useful in the semiclassical thermodynamic limit \cite{Gromov:2011jh,Kostov:2012yq,Kostov:2012jr,Bettelheim:2014gma,Kostov:2014iva}. An interesting question is whether the one-point overlap~(\ref{overlap})  also admits a determinant representation.

In this paper we investigate this question in the simplest case when the auxiliary space has dimension two ($k=2$). We  have found that the answer  is affirmative, and moreover the result is given by exactly the same determinant formula as the overlap with the N\'eel state  \cite{Brockmann:2014a,Brockmann:2014b},  upon relaxing the half-filling condition $M=L/2$ necessary to make the N\'eel overlap non-zero. 
The final result is written in terms of the matrices of size
$M/2\times M/2$: 
\begin{equation}\label{Kpm}
 K^\pm_{jk}=\frac{2}{1+\left(u_j-u_k\right)^2}\pm
 \frac{2}{1+\left(u_j+u_k\right)^2}\, ,
\end{equation}
and
\begin{equation}\label{Gpm}
 G^\pm_{jk}=\left(\frac{L}{u_j^2+\frac{1}{4}}-\sum_{n}^{}K^+_{jn}\right)\delta _{jk}
 +K^\pm_{jk}.
\end{equation}
The structure constant (\ref{overlap}) is given by the ratio of two determinants:
\begin{equation}\label{C-overlap}
 C_{\mathbf{u}}=2\left[
 \left(\frac{2\pi ^2}{\lambda }\right)^L\,\frac{1}{L}\,
 \prod_{j}^{}\frac{u_j^2+\frac{1}{4}}{u_j^2}\,\,\frac{\det G^+}{\det G^-}\right]^{\frac{1}{2}}.
\end{equation}

When $M=L/2$, this formula coincides exactly with the expression for overlap between a half-filled Bethe eigenstate and the N\'eel state  given in   \cite{Brockmann:2014a,Brockmann:2014b}. Although the MPS is different from the N\'eel state, even if restricted to  equal number of up and down spins, this is not a coincidence. We were able to show that the MPS is cohomologically equivalent to the N\'eel state at  half filling and consequently has the same overlaps with all half-filled Bethe eigenstates. The result above then follows from the derivation in \cite{Pozsgay:2009,Brockmann:2014a,Brockmann:2014b}, for $M=L/2$. When $M<L/2$, this formula is a conjecture which we have extensively checked. We have also identified a natural generalization of the N\'eel state away from half-filling, which lies in the same cohomology class as the definite-spin projection of MPS.

In section~\ref{calculation} we introduce the tools necessary for our computation, namely the Bethe ansatz and 
an explicit realization  of a set of $k\times k$ matrices which constitute a unitary $k$-dimensional represen\-ta\-tion 
of $SU(2)$. Subsequently, in section 4 we sketch  our 
computations and present the results. In section 5 we discuss the relationship between the MPS and the N\'eel state and introduce generalized N\'eel states with unequal number of up and down spins.
Section 6 contains a discussion of the thermodynamical limit and section 7 some comments on 
the string theory observables dual
to the one-point functions of the defect CFT. Finally section 8 contains our conclusion.

\section{Setting up  the computation \label{calculation}}

Although the construction of the defect state has a strong resemblance with certain 
elements of the algebraic Bethe ansatz we have found it most convenient to evaluate the overlaps by using the Bethe ansatz in its coordinate space version which we will summarize below, see for instance \cite{Bethe:1931hc,Karbach:1998}.
Hereafter we will present the explicit
representations of $SU(2)$ that we will make use of in our computations.

\subsection{The coordinate Bethe Ansatz} 

The eigenstates of the dilatation operator restricted to the $SU(2)$ sector are in one-to-one correspondence with eigenstates of the Heisenberg XXX spin chain. In this section we introduce this model and discuss its solution via the coordinate Bethe ansatz.

\paragraph{Model}The XXX spin chain is a one-dimensional lattice model consisting of $L$ spin-$\half$ particles. Therefore, the Hilbert space is $\bigotimes_L \mathbb{C}^2$, where each $\mathbb{C}^2$ is spanned by $ |\! \uparrow\rangle, |\! \downarrow\rangle$. The Hamiltonian describes a standard nearest neighbor spin-spin interaction
\begin{align}\label{Heis}
&\mathcal{H} = \sum_{i=1}^{L} \mathcal{H}_{ii+1},
&&\mathcal{H}_{ij} = \frac{1}{4} - \vec{S}_i\cdot\vec{S}_{j},
\end{align}
with periodic boundary conditions $L+1\equiv 1$. For simplicity let us also introduce the the usual raising and lowering operators $S^\pm$ such that
\begin{align}
&S^+ |\! \downarrow\rangle = |\! \uparrow\rangle,
&&S^- |\! \uparrow\rangle = |\! \downarrow\rangle.
\end{align}
Expressing the permutation operator in terms of spin operators one can see that (\ref{dilop}) reduces to (\ref{Heis}) in the $SU(2)$ subsector, up to normalization.
The (coordinate) Bethe ansatz gives us a method to diagonalize this Hamiltonian and to compute its spectrum.

\paragraph{Bethe eigenstates} The first step of the Bethe ansatz is to introduce a vacuum state
\begin{align}\label{eq:vacuum}
|0\rangle = \bigotimes_{i=1}^L | \! \uparrow \rangle.
\end{align}
This vacuum state is trivially an eigenstate of the Hamiltonian. The other eigenstates will also have down-spins on various sites. The Bethe ansatz postulates that these eigenstates are of a plane wave type. More precisely, each flipped spin behaves like a quasi-particle referred to as a magnon. These magnons propagate along the spin chain with some definite momentum $p$. The Bethe eigenstate for a chain of length $L$ describing $M$ magnons, is of the form
\begin{align}\label{eq:GenEigenstate}
|\vec{p}\rangle := |p_1,\ldots, p_M\rangle = N \sum_{\sigma \in S_{M}} \sum_{1\leq n_1<\ldots<n_M\leq L} e^{i
\sum_m (p_{\sigma_m} n_m + \sum_{j<m} \frac{\theta_{\sigma_j\sigma_m}}{2} )} S^-_{n_1}\ldots S^-_{n_M} |0\rangle,
\end{align}
where $N$ is an overall normalization. The sum over $\sigma$ runs over all permutations of $M$ elements. Furthermore, the factors $\theta$ parameterize the two-magnon S-matrix via
\begin{align}
\mathcal{S}_{ij}  := e^{\theta_{ij} - \theta_{ji}} = -\frac{1+ e^{i p_i + i p_j} - 2 e^{ip_i}}{1+ e^{i p_i + i p_j} - 2 e^{ip_j}}.
\end{align}
It is worthwhile to note that, up to an overall normalization, the Bethe vector \eqref{eq:GenEigenstate} only depends on the S-matrix $\mathcal{S}$ rather than the phase $\theta$. In the remainder we will choose the normalization $N$ such that the term $e^{i p_i n_i}$ (\textit{i.e.} the term with $\sigma=1$) in \eqref{eq:GenEigenstate} appears with unit coefficient. In other words, we will set $N = e^{-\sum_{j<k} \theta_{jk}/2}$.

\paragraph{Bethe equations} Finally, the state \eqref{eq:GenEigenstate} should respect the correct boundary conditions, i.e.\ it should be periodic. Imposing periodicity results in a set of equations on the momenta of the magnons, called the Bethe equations
\begin{align}\label{eq:BAE}
e^{ip_k L}  = \prod_{i\neq k} \mathcal{S}_{ki}.
\end{align}
When the momenta satisfy these Bethe ansatz equations, it is easy to check that the state \eqref{eq:GenEigenstate} is an eigenstate of the Hamiltonian with eigenvalue
\begin{align}
E = 2 \sum_{i=1}^M \sin^2 \frac{p_i}{2}=\frac{1}{2}\sum_{i=1}^M \frac{1}{u_i^2+\frac{1}{4}},
\end{align}
where  $u=\frac{1}{2}\cot(p/2)$ is the rapidity.
In order for a Bethe eigenstate to represent a single trace gauge theory operator it is furthermore necessary that the momenta
of its excitations add up to an integer multiple of $2\pi $. This is required to account for the cyclicity properties of the trace, i.e.
\begin{align}
P\equiv \sum_{i=1}^M p_i=2\pi m.
\end{align}
Finally, notice that our Bethe states \eqref{eq:GenEigenstate}  (with $N = e^{-\sum_{j<k} \theta_{jk}/2}$) are not normalized to unity. 
These coordinate space Bethe eigenstates can be related to the eigenstates of the algebraic Bethe ansatz approach in the
following way (see, for example, \cite{Escobedo:2010xs})
\begin{align}\label{eq:NormABA}
|\{u_i\}\rangle&=B(u_1)\ldots B(u_M)|0\rangle  \nonumber\\
&=\prod_{j}\left(u_j-\frac{i}{2}\right)^L \left( \frac{i}{u_j+\frac{i}{2}}\right)
\prod_{l<m} \left(1+\frac{i}{u_l-u_m}\right)
|p_1,\ldots, p_M\rangle.
\end{align}
This, in conjunction with the Gaudin formula \cite{Gaudin:1976sv,Korepin:1982gg} for the norm of $|\{u_i\}\rangle$, fixes the normalization of coordinate Bethe ansatz eigenstates.

\paragraph{Overlap}

Let us now continue by computing the overlap between the Bethe states and the defect state  $\langle{\rm MPS}\,|\vec{p}\rangle$. Inserting the $M$-magnon state \eqref{eq:GenEigenstate} into \eqref{eq:overlap} yields
\begin{align}\label{eq:OverlapM2}
\langle {\rm MPS}\,|\vec{p}\rangle = N \sum_{\sigma \in S_{M}} \sum_{1\leq n_1<\ldots<n_M\leq L} e^{ip_{\sigma(i)} n_i + \sum_{j<i} \ihalf\theta_{\sigma(j)\sigma(i)}} \, \mathrm{tr}[ t_1^{n_1-1}t_2t_1^{n_2-n_1-1}\ldots],
\end{align}
where the $t_i$ form the standard $k$-dimensional irreducible representation of $\mathfrak{su}(2)$. However, for practical computations it is more convenient to take
\begin{align}\label{eq:OverlapM}
\langle {\rm MPS}\,|\vec{p}\rangle =N \sum_{\sigma \in S_{M}} \sum_{1\leq n_1<\ldots<n_M\leq L} e^{ip_{\sigma(i)} n_i + \sum_{j<i} \ihalf\theta_{\sigma(j)\sigma(i)}} \, \mathrm{tr}[ t_3^{n_1-1}t_1t_3^{n_2-n_1-1}\ldots],
\end{align}
which will clearly yield the same results.

\subsection{Representations of $\mathfrak{su}(2)$}

Let us spell out the explicit representation for the $\mathfrak{su}(2)$ generators $t_i$ that we will use and derive some useful relations.

\paragraph{Definition} 
Consider the $k$-dimensional complex vector space generated by the basis vectors $E_i$. Define the standard matrix unities $E^i{}_j$ that are zero everywhere except for a 1 at position $(i,j)$, such that they satisfy
\begin{align}
E^i{}_{j}E^k{}_{l} = \delta^k{}_j E^i{}_l.
\end{align}
If we introduce the following constants
\begin{align}
&c_{k,i} = \sqrt{i(k-i)},
&& d_{k,i} = \half(k-2i+1),
\end{align}
and consider the matrices
\begin{align}\label{eq:tgenerators}
& t_+:= \sum_{i=1}^{k-1} c_{k,i} E^i{}_{i+1},
&& t_- :=\sum_{i=1}^{k-1}  c_{k,i} E^{i+1}{}_{i},
&& t_3 :=\sum_{i=1}^k  d_{k,i} E^{i}{}_{i}, 
\end{align}
then we obtain the standard $k$-dimensional $\mathfrak{su}(2)$ representation by defining
\begin{align}\label{eq:defrep}
&t_1 = \frac{t_+ +t_-}{2},
&&t_2 = \frac{t_+ -t_-}{2i}.
\end{align}
It is easy to check that these matrices satisfy the $\mathfrak{su}(2)$ commutation relations \eqref{eq:su2relations}. Note that all these matrices are traceless.

\paragraph{Automorphisms}

Let us introduce two similarity transformations
\begin{align}\label{eq:autSU2}
&U = U^{-1} := \sum_{i=1}^k E^i{}_{k-i+1},
&&V = V^{-1} := \sum_{i=1}^k (-1)^i E^i{}_{i}.
\end{align}
It is easy to show that under these transformations
\begin{align}
&U t_1 U^{-1} = t_1, &&U t_{2,3} U^{-1} = -t_{2,3}
&&V t_3 V^{-1} = t_3, &&V t_{1,2} V^{-1} = -t_{1,2}.
\end{align}
Hence, they provide a trivial automorphism of the algebra.

\section{Results \label{results}}

In this section we present a number of explicit results for the overlap \eqref{eq:OverlapM}.

\subsection{$L$ or $M$ odd} If $L$ or $M$ is odd,  the overlap vanishes. This follows directly from the automorphisms \eqref{eq:autSU2}. Indeed, for any state of the form $\mathrm{tr} [t_3^{n_1} t_1 t_3^{n_2}\ldots]$, containing $M$ $t_1$'s and $L$ $t$'s we have by cyclicity of the trace
\begin{align}
\mathrm{tr} [t_3^{n_1} t_1 t_3^{n_2}\ldots] =
\mathrm{tr} [(Ut_3U^{-1})^{n_1} Ut_1U^{-1} (Ut_3U^{-1})^{n_2}\ldots]  =
(-1)^{L-M} \mathrm{tr} [t_3^{n_1} \ldots] 
\end{align}
and
\begin{align}
\mathrm{tr} [t_3^{n_1} t_1 t_3^{n_2}\ldots] = \mathrm{tr} [(Vt_3V^{-1})^{n_1} Vt_1V^{-1} (Vt_3V^{-1})^{n_2}\ldots]  =
(-1)^{M} \mathrm{tr} [t_3^{n_1} \ldots].
\end{align}
This implies that the expression $\mathrm{tr} [t_3^{n_1} t_1 t_3^{n_2}\ldots]$, and hence the overlap \eqref{eq:OverlapM}, is only non-vanishing if $L$ and $M$ are both even.

\subsection{Vacuum, $M=0$}
From \eqref{eq:tgenerators} we see that $t_3$ is a diagonal matrix with entries $\half(k-2i+1)$ for $i=1,\ldots,{k}$. From this, it immediately follows that for the vacuum state \eqref{eq:vacuum} the overlap \eqref{eq:OverlapM} reduces to
\begin{align}
\langle{\rm MPS}\,|0\rangle =  \mathrm{tr}\, t_3^L = \sum_{i=1}^{k} d_{k,i}^L.
\end{align}
The resulting sum can be evaluated to a combination of $\zeta$-functions
\begin{align}\label{eq:M0overlap}
\langle{\rm MPS}\,|0\rangle =  \zeta_{-L}(\sfrac{1-k}{2})- \zeta_{-L}(\sfrac{1+k}{2}).
\end{align}
Taking the $k\rightarrow\infty$ limit of the explicit expression for $\langle{\rm MPS}\,|0\rangle$ yields
\begin{align}\label{eq:largekvacuum}
\langle{\rm MPS}\,|0\rangle =  \frac{k^{L+1}}{2^L(L+1) } + \mathcal{O}(k^L)\qquad \left(k\rightarrow \infty \right).
\end{align}
This agrees with the large $k$ behavior which was found previously in \cite{Nagasaki:2012re,Kristjansen:2012tn}. 

\subsection{Excited states}

\subsubsection{General considerations\label{generalities}}

We first notice that the defect state $\left|{\rm MPS}\rangle\right.$ is a cyclically invariant state (due to the cyclic nature of 
its expansion coefficients).  This implies that 
\begin{align}
\left(\left\langle {\rm MPS}\right.\left| \vphantom{{\rm MPS}}\,U\right)\,\right|\left. \vphantom{{\rm MPS}}\vec{p}\right\rangle=
\left\langle {\rm MPS}\right.\left| \vphantom{{\rm MPS}}\right.\left.\vphantom{{\rm MPS}} \vec{p}\right\rangle=
\left\langle {\rm MPS}\right.\left|\vphantom{\Psi ^{\rm cl}} \,\left(\vphantom{{\rm MPS}}U\,\right|\left. \vec{p}\vphantom{{\rm MPS}}\right\rangle\right),
\end{align}
where $U=e^{i\hat{P}}$ is the lattice translation operator and $\hat{P}$ the momentum operator.
 From this we conclude that the overlap vanishes unless 
$\left| \vec{p}\right\rangle$ is a zero-momentum state.

Secondly, we notice that for an even number of excitations 
$\left|{\rm MPS}\rangle\right.$ is invariant under an operation traditionally denoted as parity, see for instance~\cite{Beisert:2003tq}.
Its
action on a spin state is defined by
\begin{align}
{\cal P} \left.\left| {\mathfrak t}_1 {\mathfrak t}_2\ldots {\mathfrak t}_n \right.  \right \rangle=
 \left.\left| {\mathfrak t}_n {\mathfrak t}_{n-1}\ldots {\mathfrak t}_1 \right.  \right \rangle,
\end{align}
where ${\mathfrak t}_i\in \{ \downarrow, \uparrow \}$.
The invariance of  $\left|{\rm MPS}\rangle\right.$under this transformation follows from the invariance of its expansion
coefficients under a similar operation performed on the matrices inside the traces.  By an argument similar to the one
above it follows that the overlap vanishes unless the Bethe eigenstate has positive parity. It is well-known that the 
eigenstates of the Heisenberg spin chain can be chosen to be eigenstates of a definite parity. In particular, the 
so-called un-paired eigenstates for which the Bethe rapidities fulfill that $\{u_i\}=\{-u_i\}$ are automatically eigenstates with parity equal to 
$(-1)^{M(L+1)}$. Moreover, as discussed in section \ref{sec:Defect}, we find that only these unpaired state can have a non-trivial overlap with the classical function. This follows from the fact that the unpaired states are exactly the states that are annihilated by the odd charges $Q_{2n+1}$. 

\subsubsection{Two excitations, $M=2$\label{two excitations}} 

By using the cyclicity of the trace, we can rewrite the overlap \eqref{eq:OverlapM} as a sum of terms of the form
\begin{align}
\mathrm{tr} [ t_3^{L-m-1} t_1 t_3^{m-1} t_1 ].
\end{align}
We can evaluate this trace by implementing the explicit expressions for $t_i$ \eqref{eq:tgenerators}
\begin{align}
\mathrm{tr} [ t_3^{L-m-1} t_1 t_3^{m-1} t_1 ] = &\sum_{a,b=1}^k\sum_{i,j=1}^{k-1}
\frac{1}{4}d_{k,a}^{L-m-1}d_{k,b}^{m-1}c_{k,i}c_{k,j}\nonumber\\
&\qquad \mathrm{tr} [ 
 E^a{}_a 
\left(E^i{}_{i+1} + E^{i+1}{}_{i} \right)
E^b{}_b 
\left(E^j{}_{j+1} + E^{j+1}{}_{j} \right)
].
\end{align}
The definition of the matrix unities then allows us to work out the trace
\begin{align}
\mathrm{tr} [ t_3^{L-m-1} t_1 t_3^{m-1} t_1 ] 
& = 
2^{1-L}\sum_{i=1}^{k-1}
\frac{i(k-i)}{(k-2i)^2 - 1}
\left[\frac{k-2i+1}{k-2i-1}\right]^m (k-2i-1)^L.
\end{align}
%
Thus, for $M=2$, the Bethe states are mapped to
\begin{align}
\langle{\rm MPS}\, | p_1,p_2\rangle &= \sum_{m<n} [e^{i (p_1 n+p_2 m)} + \mathcal{S}_{21} e^{i (p_2 n+p_1 m)} ] \mathrm{tr} [ t_3^{m-1} t_1 t_3^{n-m-1} t_1 t_3^{L-n} ] \nonumber\\
&= \sum_{m<n} [e^{i (p_1 n+p_2 m)} + \mathcal{S}_{21} e^{i (p_2 n+p_1 m)} ] \mathrm{tr} [t_3^{L-n+m-1} t_1 t_3^{n-m-1} t_1].
\end{align}
The sums over $m,n$ can easily be done and we find the following formula for the overlap
\begin{align}
&\langle{\rm MPS}\, | p_1,p_2\rangle =
\frac{e^{i(p_1+p_2)}}{1-e^{i(p_1+p_2)}}\sum_{i=1}^{k-1}\frac{i(k-i)}{2^{L-1}(k-2i-1)^{2-L}} \left[e^{ip_2} \frac{e^{iLp_2} \left[\frac{k-2i+1}{k-2i-1}\right]^L -1}{e^{ip_2} \left[\frac{k-2i+1}{k-2i-1}\right] -1} - \right.\\
&\left. 
e^{iLp_2} \frac{e^{iLp_1} - \left[\frac{k-2i+1}{k-2i-1}\right]^L }{e^{ip_1} - \left[\frac{k-2i+1}{k-2i-1}\right] } 
+\mathcal{S}_{21} e^{ip_1} \frac{e^{iLp_1}  \left[\frac{k-2i+1}{k-2i-1}\right]^L - 1}{e^{ip_1} \left[\frac{k-2i+1}{k-2i-1}\right] -1 } 
- \mathcal{S}_{21} e^{iLp_1} \frac{e^{iLp_2} - \left[\frac{k-2i+1}{k-2i-1}\right]^L }{e^{ip_2} - \left[\frac{k-2i+1}{k-2i-1}\right] } 
\right]. \nonumber
\end{align}
Notice that the above expression has to be evaluated with care in case $k$ is odd due to a superficial pole at $i=\half(k-1)$. By using that $\langle{\rm MPS} \,| p_1,p_2\rangle $ is invariant if we redefine the summation via $i\rightarrow k-i$ it is easy to check that upon substituting the Bethe equations \eqref{eq:BAE} the overlap vanishes unless
$p_1+p_2=0$ where the above expression has a pole.  Then, imposing the vanishing of the total momentum and setting $p_1=-p_2=p$ from the beginning gives us the following one-point function
\begin{align}
\langle {\rm MPS}\,| p,-p\rangle = L u (u-\ihalf) \sum_{j=-\frac{k}{2}}^{\frac{k}{2}} \frac{j^2 - \frac{k^2}{4}}{j^2 + u^2}(j-\half)^{L-1}.
\end{align}
For $k=2$ this reduces to $2^{1-L}\, L u^{-1} (u-\ihalf)$.

\subsubsection{General $M$}

In the following we will  derive some results for a general even number of excitations $M$. 
In particular, for the case $k=2$, we will give a closed formula of determinant form, valid for any even $M$.

\paragraph{$k=2$} For $k=2$ computing the overlap simplifies due to the identities
\begin{align}
&t_i^2 = \sfrac{1}{4},
&& \{ t_i , t_j \} = 0, \qquad i\neq j.
\end{align}
The anti-commutator identity means that we can order the generators in the trace (possibly at the cost of a sign) and the first identity implies that we can take all the powers in the trace mod 2. In particular, we can simplify \eqref{eq:OverlapM} to
\begin{align}
\langle{\rm MPS}\,|\vec{p}\rangle_{k=2} &= N \sum_{\sigma \in S_{M}} \sum_{1\leq n_1<\ldots<n_M\leq L} e^{ip_{\sigma(i)} n_i + \sum_{j<i} \ihalf\theta_{\sigma(j)\sigma(i)}} \, (-1)^{\sum_i n_i +\frac{M}{2}}\mathrm{tr}[ t_1^{L-M}t_2^M], \nonumber\\
&= \frac{(-1)^{M/2} N}{2^L} \sum_{\sigma \in S_{M}} \sum_{1\leq n_1<\ldots<n_M\leq L} e^{i(p_{\sigma(i)}+\pi ) n_i + \sum_{j<i} \ihalf\theta_{\sigma(j)\sigma(i)}} ,\nonumber\\
&= \frac{(-1)^{M/2} N}{2^L} \sum_{\sigma \in S_{M}} e^{ \sum_{j<i} \ihalf\theta_{\sigma(j)\sigma(i)}} \sum_{1\leq n_1<\ldots<n_M\leq L} e^{i(p_{\sigma(i)}+\pi ) n_i } .
\end{align}
The above sum can be evaluated as follows
\begin{align}
&\sum_{1\leq n_1 < \ldots < n_M \leq L}  x_1^{n_1} \ldots x_M^{n_M} =  \\ 
& \hspace{2cm}
\prod_{n=1}^M x_n^{L+1} + \sum_{a=1}^M 
\!\left[\! 1-\prod_{n=1}^a x_n^{L+1} \! \right]\!
\!\left[\!\prod_{m=1}^a \frac{x_m^{m}}{1-\prod_{n=m}^a x_n} \! \right]\!
\!\left[\! \prod_{m=a+1}^M \frac{x_m^{L+1}}{\prod_{n=a+1}^m x_n-1} \! \right] .  \nonumber
\end{align}
In agreement with our general discussion, cf.\  section~\ref{sec:Defect}, we find that the only 
Bethe eigenstates that give a non-zero overlap function are states with momentum configurations of the form
\begin{align}
(p_1,-p_1,p_2,-p_2,\ldots,p_{\frac{M}{2}},-p_{\frac{M}{2}}).
\end{align}
For these states one can write the overlap function in a compact form as the determinant of a matrix. Define the following function
\begin{align}
K_{ij} : = \frac{1}{2} \left[ \frac{1 + 4 u_i^2} {1 + (u_i + u_j)^2} + \frac{1 + 4 u_i^2} {1 + (u_i - u_j)^2}  \right],
\end{align}
and the following $M/2\times M/2$ matrix
\begin{align}
A_{ij} : = (L - \sum_{n=1}^{M/2} K_{in}) \delta_{ij} + K_{ij},
\end{align}
then the overlap function is given by
\begin{align}
\langle{\rm MPS}\,|\vec{p}\rangle_{k=2} = 2^{1-L}
(\det A) \prod_{i=1}^{M/2} \frac{u_i - \ihalf}{u_i}.
\end{align}
We have confirmed this formula by explicit computations up to and including the case of eight excitations. Upon translating to the algebraic Bethe ansatz framework (cf. \eqref{eq:NormABA}), using the Gaudin formula for the norm, and applying elementary determinant identities, we arrive at the aforementioned result \eqref{C-overlap}.

\paragraph{Large $k$} Let us have a closer look at the leading order large $k$ expansion for any number of excitations. One can show that for $M$ excitations
\begin{align}\label{eq:largek}
\mathrm{tr} ( t_3^{n_1-1} t_1 t_3^{n_2-n_1-1} t_1\ldots )  =-
\frac{
\sqrt{\pi} \, \Gamma \left( -\frac{L+1}{2} \right)}{\Gamma
\left(\frac{1-M}{2}\right) \Gamma \left(\frac{ 1-L+M}{2} \right)} k^{L+1}  + \mathcal{O}(k^L).
\end{align}
This can be seen as follows. First, in the large $k$ limit $\mathrm{tr} ( t_3^{L-M} (t_+ t_-)^{\frac{M}{2}}) $ can be rewritten as a Riemann sum and integration then  leads to the following identity
\begin{align}\label{eq:orderlargek}
\mathrm{tr} ( t_3^{L-M} (t_+ t_-)^{\frac{M}{2}})  =-
 \frac{
\sqrt{\pi} \, \Gamma \left( -\frac{L+1}{2} \right)}{\binom{M}{\frac{M}{2}}\Gamma
\left(\frac{1-M}{2}\right) \Gamma \left(\frac{ 1-L+M}{2} \right)} \frac{k^{L+1}}{\binom{M}{\frac{M}{2}} 2^{L+1}}  + \mathcal{O}(k^L).
\end{align}
Second, from the defining commutation relations of $\mathfrak{su}(2)$ it can be seen that any distribution of $t_3,t_\pm$ under the trace can be ordered as \eqref{eq:orderlargek} at the cost of terms of lower order in $k$. Then \eqref{eq:largek} follows by expressing $t_1$ in terms of $t_\pm$ as in \eqref{eq:defrep}.

This means that the large $k$ limit of the overlap function reduces to
\begin{align}\label{eq:leadingkM}
\langle {\rm MPS}\,| \vec{p} \rangle =
-\sqrt{\pi}
\frac{N\,
\Gamma \left( -\frac{L+1}{2} \right)}{\Gamma
\left(\frac{1-M}{2}\right) \Gamma \left(\frac{ 1-L+M}{2} \right)} \frac{k^{L+1}}{2^L}
\sum_{\sigma \in S_{M}} \sum_{1\leq n_1<\ldots<n_M\leq L} e^{ip_{\sigma(i)} n_i + \sum_{j<i} \ihalf\theta_{\sigma(j)\sigma(i)}}.
\end{align}
It is easy to check that for $M=0$ it reduces to the large $k$ behavior we found for the vacuum state \eqref{eq:largekvacuum}. However, for $M\neq 0$ something unusual happens.

Notice that \eqref{eq:leadingkM} can be expressed as the inner product of the Bethe state \eqref{eq:GenEigenstate} with the fully symmetrized state that has $M$ spins down. Such a state can be expressed as the lowering operator $S^-$ acting on vacuum $M$ times, i.e. $\Delta^{(L)}(S^-)^M | 0 \rangle$. Thus, we can re-express the overlap as
\begin{align}
\langle {\rm MPS}\, | \vec{p} \rangle =
\langle 0 | \Delta^{(L)} (S^+)^M | \vec{p}\rangle,
\end{align}
where $\Delta$ is the coproduct. However, due to the fact that Bethe states are highest weight states, the above vanishes. In other words, the inclusion of excitations lowers the order of the overlap for large $k$.

In order to gain a better understanding of this phenomenon, let us look at the large $k$ behavior for $M=0,2,4$. We study the large $k$ behavior by explicitly evaluating the relevant overlap function for a large range of values of $L,k$. The overlap will be a polynomial in $k$ of degree at most $L+1$ with coefficients that are rational functions of $L$. Letting $L$ run from 2 to $20$ and $k$ from $2$ to $30$ allowed us to fix the relevant coefficients. In general, we find that the large $k$ behavior is of the form
\begin{align}\label{eq:generalKbig}
\langle {\rm MPS}\, | \vec{p} \rangle =
&N \sum_{\sigma} \sum_{n_i}     
\, \sum_{m=0} \beta^{(m)}_{L,M}(n_i) k^{L+1-m} \,
e^{ip_{\sigma(i)} n_i + \sum_{j<i} \ihalf\theta_{\sigma(j)\sigma(i)}}.
\end{align}
The coefficient $\beta^{(0)}$ is constant and can be read off from \eqref{eq:leadingkM}. For $M=0$ the first few $\beta^{(m)}$ are constant and from \eqref{eq:M0overlap} the large $k$ behavior is easily found to be
\begin{align}
\langle{\rm MPS}\,|0\rangle = \frac{1}{2^L} \left(
\frac{k^{L+1}}{L+1}  -\frac{1}{6} L \,k^{L-1} + 
\frac{7}{360} (L-2) (L-1) L\, k^{L-3}
+ \mathcal{O}(k^{L-5})\right).
\end{align}
Notice that the even orders vanish.

However, starting from $M=2$ the coefficients become non-trivial. Let us list the first few $\beta^{(m)}_{L,2}$ and describe their contribution. If we denote $n_{ij} = n_i - n_j$, then
\begin{align}
& \beta^{(1)}_{L,2} = \frac{2^{-L} }{L-1}, \\
& \beta^{(2)}_{L,2} = \frac{2^{1-L} }{L-3} \left[ \frac{L}{3}+ \frac{n_{12}(L+n_{12})}{L-1}\right] \\
& \beta^{(3)}_{L,2} =  \frac{L(L+1) + 6\, n_{12}(L+n_{12})}{3\cdot 2^L (L -3)}\\
&\ \beta^{(4)}_{L,2} = \frac{2^{1-L}}{L-5} \left[\frac{ (L-2) L (L+3)}{30} + \frac{(L^2-4 L+5) n_{12} (L+n_{12})}{3(L-3)}+\frac{n_{12}^2 (L+n_{12})^2}{3(L-3)}\right]
\end{align}
Since $\beta^{(1)}_{L,2} $ is constant it vanishes by the same arguments as the leading order. For the other terms, the factors of $n_i$ can be written as derivatives of momenta $p_i$ when calculating the explicit overlap function. This allows us to evaluate the overlap \eqref{eq:generalKbig} to the relevant order. Again we find that upon using the Bethe equations that it vanishes unless we impose pairwise momentum conservation. Doing this, we find for the next two terms
\begin{align}
\langle {\rm MPS}\,|p,-p\rangle & = \frac{u(u+\ihalf) L }{L-3} \left[ \frac{k^{L-1}}{2^{L-2}} +(L-1) \frac{k^{L-2}}{2^{L-2}}  +  \mathcal{O}(k^{L-3}) \right]
\end{align}
Notice that, in contradistinction to the vacuum, there is a contribution at an even order. Finally, the next non-trivial contribution is
\begin{align}
\langle {\rm MPS}\,|p,-p\rangle_{\mathcal{O}(k^{L-3})} =& 
\frac{2^{2-L}L(L-1)}{3(L-3)(L-5)} u(u+\ihalf)[L(L-11) -12 u^2].
\end{align}
Starting from $k^{L-1}$ terms appear at both even and odd orders. 

Next, we turn to four excitations $M=4$. It can be shown that the first order for $M=4$ particles that contributes is $k^{L-3}$. This seems to indicate that the order at which the large $k$ expansion begins is $k^{L-M+1}$. The first non-trivial coefficient for four particles can be computed along the same lines as for $M=2$ and we find
\begin{align}
\frac{u_1(u_1+\ihalf)u_2(u_2+\ihalf)}{2^{L-4}}\frac{L}{L-7}\left[L-4 + \frac{2(1+u_2^4 + u_1^2(1-8u_2^2))}{(1+(u_1+u_2)^2)(1+(u_1-u_2)^2)}\right] \, k^{L-3}
\end{align}
The general structure of the contributions is indicated in Table \ref{tab1}.

\begin{table}
\begin{center}
\begin{tabular}{c|c|c|c|c|c|}
  & $k^{L+1}$ & $k^{L}$ & $k^{L-1}$ & $k^{L-2}$ & $k^{L-3}$ \\ \hline
$M=0$ & $\star$ & 0& $\star$ & 0& $\star$ \\ \hline
$M=2$ & 0 & 0 & $\star$ & $\star $& $\star$ \\ \hline
$M=4$ & 0 & 0 & 0 & 0& $\star$ \\ \hline
\end{tabular}
\end{center}
\caption{Large $k$ behavior of the one-point functions for $M=0,2,4$ excitations. The order at which the expansion starts is $k^{L+1-M}$.}\label{tab1}
\end{table}

\section{Matrix product and N\'{e}el states}

In this section we elucidate the relationship between the matrix product and the 
N\'{e}el states. This will allow us to prove equation \eqref{C-overlap} for $M=L/2$. The N\'{e}el state is the vacuum of the classical (Ising) anti-ferromagnet:
\begin{equation}
 \left|\neel\right\rangle=\left|\ur\dr\ur\dr\ldots \ur\dr\right\rangle
 +\left|\dr\ur\dr\ur\ldots \dr\ur\right\rangle.
\end{equation}
The state has equal number of up and down spins (we assume that the length $L$ of the spin chain is even). 

On the other hand, the matrix product state has components with any even number of up and down spins. Since the total spin in conserved, it is convenient to decompose this state into components with definite number of up and down spins. Let us denote the projector onto states with $M$ down spins by $P_M$, and select the definite-spin component of the MPS (\ref{MPS}) by
\begin{equation}
 \left|\mps_M\right\rangle=P_M\left|{\rm MPS}\right\rangle.
\end{equation}
To facilitate the bookkeeping, it is convenient to introduce the generalized MPS:
\begin{equation}
 \left|\mps(z)\right\rangle=\mathop{\mathrm{tr}}_a\prod_{l=1}^{L}
 \left(t_1\left|\ur_l\right\rangle+zt_2\left|\dr_l\right\rangle\right)
\end{equation}
where $z$ is a complex number.
Then,
\begin{equation}
 \left|\mps_M\right\rangle=\oint\frac{dz}{2\pi iz^{M+1}}\,\left|\mps(z)\right\rangle.
\end{equation}

We can also generalize the N\'{e}el state to the case of an arbitrary even number of down spins:
\begin{equation}\label{generalized-Neel}
 \left|\neel_M\right\rangle=\sum_{{{l_1<\ldots <l_M}\atop{|l_i-l_j|\,-\,{\rm even}}}}^{}\left|\ur\ldots \ur\underset{l_1}{\dr}\ur\ldots \underset{l_2}{\dr}\ldots \underset{l_M}{\dr}\ldots \ur\right\rangle.
\end{equation}
This looks like a descendant of the ground state, and would have been such, if not for the constraint that spin-flips hop by an even number of sites. Obviously,
\begin{equation}
 \left|\neel\right\rangle=\left|\neel_{\frac{L}{2}}\right\rangle.
\end{equation}

\begin{figure}[t]
\begin{center}
 \centerline{\includegraphics[width=10cm]{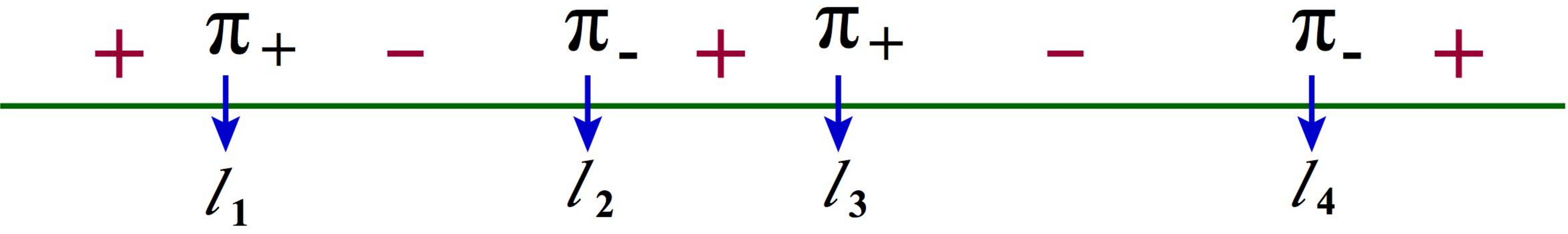}}
\caption{\label{DWall-figure}\small The generalized MPS state (\ref{domain-wall}).}
\end{center}
\end{figure}

Another state that we shall deal with is a hybrid between the generalized N\'eel and MPS\footnote{Here we assume that $m$ is even. The definition however can be extended to odd $m$, see below.}:
\begin{align}\label{domain-wall}
 \left|\mps_m(z)\right\rangle&=&
  \mathop{\mathrm{tr}}\nolimits_a
  \sum_{{{l_1<\ldots <l_m}\atop{|l_i-l_j|\,-\,{\rm even}}}}^{}
\prod_{s=1}^m 
\left[\pi_{(-)^{s+1}}\left|\dr_{l_s}\right\rangle \prod_{l=l_{s}+1}^{l_{s+1}-1}
\left(t_1\left|\ur_l\right\rangle+(-1)^s z t_2\left|\dr_l\right\rangle\right)\right],
\end{align}
where the product is understood in the cyclic sense, such that $l_{m+1}\equiv l_1$ and $l=L+k$ is identified with $l=k$.
Here $\pi _\pm$ are chiral projectors in the auxiliary space:
\begin{equation}
 \pi_\pm= \half \pm t_3.
\end{equation}
For instance, in the representation where $t_i=\sigma _i/2$, these are the ordinary spin-up/spin-down projectors:
\begin{equation}
 \pi _+=\left|\ur_a\right\rangle\left\langle \ur_a\right|,\qquad 
  \pi _-=\left|\dr_a\right\rangle\left\langle \dr_a\right|.
\end{equation}

The generalized MPS  can be pictured as a collection of $m$ domains, separated by domain walls. Each domain wall carries a down spin in the quantum space and $\pi _\pm$ projector in the auxiliary space. The sign of $z$ flips across each domain wall (fig.~\ref{DWall-figure}). Since $\pi _\pm$ are projectors, the trace over the auxiliary space decomposes onto the product of matrix elements for each of the domains. The chirality of projectors enforces the domains to contain odd number of sites each.

The definite-spin projections of the generalized MPS,
\begin{equation}\label{mpskM}
 \left|\mps_{m,M}\right\rangle=P_M\left|\mps_m(1)\right\rangle
 =\oint\frac{dz}{2\pi iz^{M-m+1}}\,\left|\mps_m(z)\right\rangle,
\end{equation}
interpolate between the definite-spin components of the MPS and the generalized N\'eel states (\ref{generalized-Neel}). Indeed,
\begin{equation}
 \left|\mps_{0,M}\right\rangle=\left|\mps_M\right\rangle,
 \qquad 
 \left|\mps_{M,M}\right\rangle=2^{M-L}\left|\neel_M\right\rangle.
\end{equation}

All these different states are related to each other, and in fact can be all expressed through the basic MPS (\ref{MPS}) by simple projection and spin-lowering operations. In particular, we will find that definite-spin components of the MPS are cohomologically equivalent to the generalized N\'{e}el states:
\begin{equation}\label{MPS->Neel}
 \left|\mps_M\right\rangle=\frac{1}{2^L(\ihalf)^{M}} \left|\neel_M\right\rangle
 +S^- \left|\ldots \right\rangle,
\end{equation}
where $S_i$ is the total spin operator, and $S^-$ is its lowering component that flips in turn all the spins in the chain with weight one. 

Since Bethe states are highest-weight:
\begin{equation}
 S^+\left|\left\{u_j\right\}\right\rangle=0,
\end{equation}
their overlaps with the MPS and the N\'eel states coincide:
\begin{equation}
 \left\langle {\rm MPS}\,\right.\!\!\left|\left\{u_1\ldots u_M\right\}\right\rangle
 =\frac{1}{2^L(\ihalf)^{M}}\,\left\langle \neel_M\right.\!\!\left|
 \left\{u_1\ldots u_M\right\}\right\rangle
\end{equation}
The determinant representation (\ref{C-overlap}) in the case of $M=L/2$ then follows from the known overlap between the Bethe states and the ordinary N\'eel state~\cite{Pozsgay:2009,Brockmann:2014a,Brockmann:2014b}. For other $M$, the overlap is given by the same equation, which we believe is a new result, that would be interesting to prove, either directly in the MPS representation or using its cohomological equivalence to the generalized N\'eel states  (\ref{generalized-Neel}).

Now we proceed to prove (\ref{MPS->Neel}). The proof rests on the following identity:
\begin{equation}\label{MPS->MPS}
 \left(i\,\frac{d}{dz}+S^-\right)^m\left|\mps(z)\right\rangle
 =m!\,\left|\mps_m(z)\right\rangle.
\end{equation}
Though not entirely obvious, this equation can be derived in a rather straightforward way.
Both   $S_-$ and $d/dz$, when acting on $\left|\mps(z)\right\rangle$, produce $l$ terms, where the $l$-th spin is flipped, in the former case with the coefficient $t_1$ and the latter case with the coefficient $t_2$. Altogether, the action of $id/dz+S_-$ creates a defect, a down spin accompanied by $t_+$, where
\begin{equation}
 t_\pm=t_1\pm it_2.
\end{equation}
Now, taking into account that
\begin{equation}
 t_\pm t_1=t_1t_\mp,\qquad t_\pm t_2=-t_2t_\mp,\qquad t_\pm^2=0,\qquad t_\pm t_\mp =\pi _\pm,
\end{equation}
we find that
$$
 t^+\prod_{l=l_i}^{l_{i+1}}
 \left(t_1\left|\ur_l\right\rangle+zt_2\left|\dr_l\right\rangle\right)t^+
 =
\begin{cases}
 0 & {\rm~ }(l_{i+1}-l_i)\,-\,{\rm odd}
\\
 \pi _+\prod\limits_{l=l_i}^{l_{i+1}}
   \left(t_1\left|\ur_l\right\rangle-zt_2\left|\dr_l\right\rangle\right)
 \pi _- & {\rm ~}(l_{i+1}-l_i)\,-\,{\rm even}
\end{cases}
$$
from which (\ref{MPS->MPS}) immediately follows.

Applying the spin projection (\ref{mpskM}) to both sides of (\ref{MPS->MPS}) we can express the generalized MPS  through the ordinary one:
\begin{equation}
 \left|\mps_{m,M}\right\rangle=
 \sum_{s=0}^{m}
 i^{m-s}
 { M-s \choose m-s}
 \,
\frac{(S^-)^{s}}{s!}\left|\mps_{M-s}\right\rangle.
\end{equation}
The cohomological equivalence of the N\'eel states and the MPS state (\ref{MPS->Neel}) is just a particular case of this relationship.

\section{Classical limit}

If the thermodynamic limit $L\rightarrow \infty $ is accompanied by populating the spin chain with a large number of low-energy magnons, such that $M/L$ and $u_j/L$ are kept fixed as  $L\rightarrow \infty $, the spin-chain states become semiclassical \cite{Sutherland:1995zz,DharShastry,Beisert:2003xu}. Oftentimes one can directly compare spin-chain results in this regime to classical string theory in $AdS_5\times S^5$ \cite{Beisert:2003xu,Frolov:2003xy}, even though the two approximations are supposed to work in the opposite range of the 't~Hooft coupling.

In the scaling limit the Bethe roots concentrate on a number of cuts in the complex plane and are characterized by the density
\begin{equation}
 \rho (x)=\frac{1}{L}\sum_{j=1}^{\frac{M}{2}}\delta \left(x-\frac{u_j}{L}\right).
\end{equation}
We are interested in symmetric configurations, due to the selection rules for the one-point function, and define the density by summing only over the right movers which constitute one half of all Bethe roots. The density satisfies an integral equation
\begin{equation}\label{alg-curve}
 2\strokedint dy\,\rho (y)\left(\frac{1}{x-y}+\frac{1}{x+y}\right)=\frac{1}{x}+2\pi n_l,
\end{equation}
 where $n_l$ are (positive) integer mode numbers, one integer for each arc of the Bethe root distribution. The general solution to these equations can be written in terms of Abelian integrals on an algebraic curve that characterizes a particular semiclassical state of the spin chain \cite{Kazakov:2004qf}.
 
We may ask how the overlap (\ref{overlap}) behaves in this scaling limit.
The non-trivial dependence of the overlap on the Bethe roots enters through the determinants of $G^\pm$, which are structurally similar to the Gaudin determinant. The thermodynamic limit of the latter was analyzed in \cite{Gromov:2011jh} with the result that the leading contribution comes from the near-diagonal matrix elements, with $|i-j|\ll L$. But in the ratio of determinants that enters the overlap formula (\ref{C-overlap}) this contribution simply cancels, because the difference between the near-diagonal matrix elements of $G^+$ and $G^-$ is of order $1/(u_j+u_k)^2\sim 1/L^2$ and vanishes in the thermodynamic limit. One may then  expect that the ratio approaches $1$, with corrections of order $1/L$. However, the situation is more subtle, and the ratio in fact approaches a finite constant value  different from one: 
\begin{equation}
 C_{\mathbf{u}}\simeq 2K\,{\rm e}\,^{\frac{1}{2}\,L\ln\frac{2\pi ^2}{\lambda }-\frac{1}{2}\,\ln L+O\left(\frac{1}{L}\right)}\qquad \left(L\rightarrow \infty \right).
\end{equation}
The coefficient $K$ is given by the ratio of functional determinants:
\begin{equation}
 K=\left(\frac{\det\mathcal{G}^+}{\det\mathcal{G}^-}\right)^{\frac{1}{2}},
\end{equation}
where
\begin{equation}\label{opGpm}
 \mathcal{G}^\pm f(x)=-\frac{\partial }{\partial x}\strokedint
 dy\,\rho (y)\left(\frac{1}{x-y}\pm\frac{1}{x+y}\right)f(y),
\end{equation}
are operators that act in the space of functions defined on the same set of arcs in the complex plane as the Bethe root density $\rho (y)$.

The residual dependence on the density of Bethe roots arises because the original discrete determinants in (\ref{C-overlap}) have a set of nearly zero modes, as already noticed in \cite{Gromov:2011jh}. These modes corresponds to vectors $f_j$ that are approximately constant on the scale $|j-k|\ll L$. For such vectors the summation can be simply replaced by integration, and the matrices $K^\pm$ and $G^\pm$ in (\ref{Kpm}), (\ref{Gpm}) become integral operators:
\begin{equation}
 \mathcal{G}^\pm f(x)=\frac{f(x)}{x^2}-\strokedint dy\,\rho (y)\left(
 \mathcal{K}^+(x,y)f(x)-\mathcal{K}^\pm(x,y)f(y)
 \right),
\end{equation}
where
\begin{equation}
 \mathcal{K}^\pm(x,y)=\frac{2}{(x-y)^2}\pm\frac{2}{(x+y)^2}\,.
\end{equation}
Using the classical Bethe equations (\ref{alg-curve}), the $\mathcal{G}^\pm$ operators can be further simplified to (\ref{opGpm}). 

Apart from a trivial kinematic factor, the structure constant $C_{\mathbf{u}}$ does not exponentiate in the thermodynamic limit. This is perhaps an indication that the limit of large $M$ and $L$, at $k=2$, is not really classical on the string side. Indeed, the natural classical limit in string theory would also involve taking $k$ large (natural scaling is $k\sim \sqrt{\lambda }$ at strong coupling \cite{Nagasaki:2012re}). We postpone a detailed study of this limit for future work, and just make a few general comments on possible comparison to string theory in the next section.

\section{Comparison to string theory}

Earlier studies of chiral primary operators have shown that one can expect an agreement between one-point functions calculated in gauge
theory and one-point functions calculated in string theory to leading order in the parameter $\lambda/k^2$ in a double scaling limit where both
$\lambda$ and $k$ are sent to infinity but the ratio $\lambda/k^2$ is kept fixed and small.  Hence, for this purpose one would
mainly be interested
in large representations. 

The calculation of one-point functions on the string theory side was previously carried out in the case of 
chiral primary operators and involves computing the fluctuation of the probe D5 brane action due to fluctuations in the background supergravity fields when a source corresponding to the operator in question is inserted on the AdS 
boundary~\cite{Nagasaki:2012re,Kristjansen:2012tn}. The computation involved is completely analogous to the 
computation of a three-point function involving a chiral primary operator and two giant gravitons~\cite{Bissi:2011dc}, and follows a general scheme of computing one point functions in the presence of a heavy probe, such as Wilson loops \cite{Berenstein:1998ij} or  the three-point function of two heavy and one light 
operators \cite{Zarembo:2010rr,Costa:2010rz}.
Performing the calculation of one-point functions involving other types of operators would require other techniques.
One type of operators one could dream of considering could be BMN operators (i.e.\ two-excitation operators, considered
in subsection~\ref{two excitations}). The string theory dual of these  were given in~\cite{Spradlin:2006wk}. Another example
could be the operator dual to a folded spinning string with two angular momenta on $S^5$.  
This operator
is characterized by its $M\sim{\cal O}(L)$ Bethe roots being distributed on two arches placed symmetrically around zero~\cite{Beisert:2003xu,Frolov:2003xy} and belongs to the class of operators which have a non-vanishing overlap with the
defect operators, cf.\ section~\ref{sec:Defect}.
Both for BMN- and spinning string types of operators, however, it appears
that the string theoretical calculation of the one-point function would be of a similar complexity as the computation of a three-point
function involving three heavy operators. 

\section{Conclusion}

We have seen a strong indication that the integrable structures underlying the duality between ${\cal N}=4$ SYM and type IIB string
theory on $AdS_5\times S^5$  leave an imprint on the correlation functions of the defect CFT derived from the D3-D5 probe-brane set-up 
with internal gauge field flux, $k$. We have concentrated our efforts on the calculation of one-point functions of non-protected operators  and 
we have proposed, for $k=2$, a closed expression of determinant form for the one-point function of Bethe eigenstates, based on explicit computations
involving states with up to eight excitations. Furthermore, for half filling we have proved the formula by relating the matrix product state to the N\'{e}el state.
Needless to say that it would be very interesting to construct a proof of the formula in the general case. 

 The
formulation of the one-point function as an overlap involving a matrix product state could indicate interesting connections to  problems
in condensed matter physics.
In addition, there are numerous other directions of investigation which could lead to further insights on the theme touched upon here.
One- and multi-point correlation
functions of defect CFT's with dual gauge field flux 
could be studied for higher values of $k$, 
to higher loop orders and for other probe brane set-ups, such as the D3-D7 case. Finally, it would obviously be very interesting if one could match any
of these quantities with quantities derived in the dual string theory picture.

There are other cases in which heavy probes create a coherent field configuration in the CFT vacuum, which at weak coupling can be studied by semiclassical methods. This is the case for the 't~Hooft loops \cite{Gomis:2009ir}, surface operators \cite{Gukov:2006jk,Drukker:2008wr}, and domain-wall defects \cite{Kapustin:1998fk,Gaiotto:2008sa}. It would be interesting to investigate the spin-chain representation of one-point functions in these cases as well.

\begin{figure}[t]
\begin{center}
\subfigure[]{
   \includegraphics[height=1.8cm] {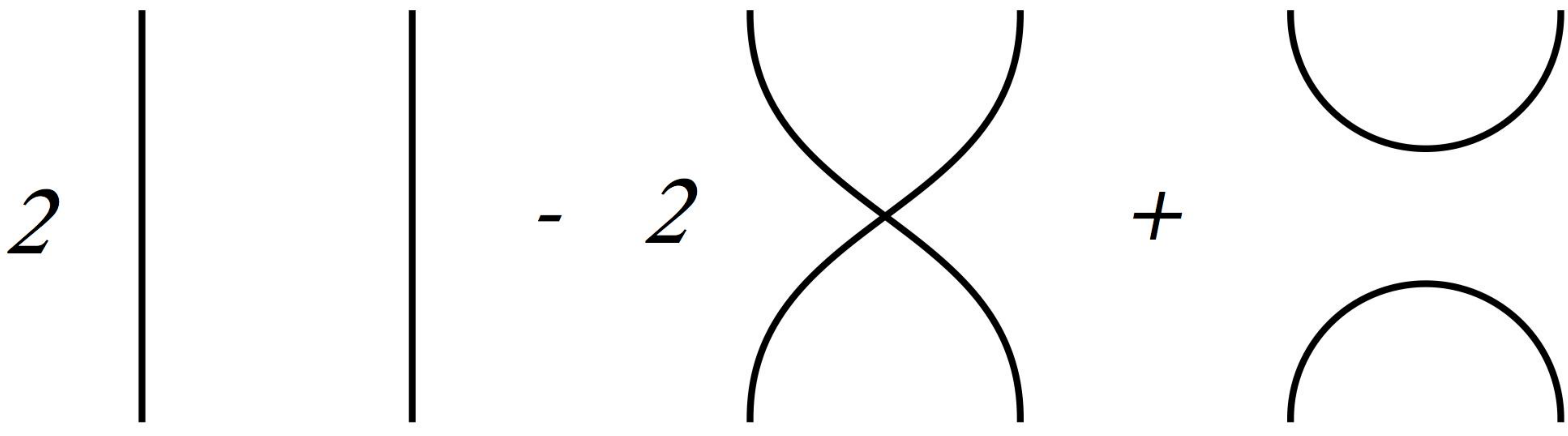}
   \label{fig:subfig1}
 }
 \subfigure[]{
   \includegraphics[height=4.5cm] {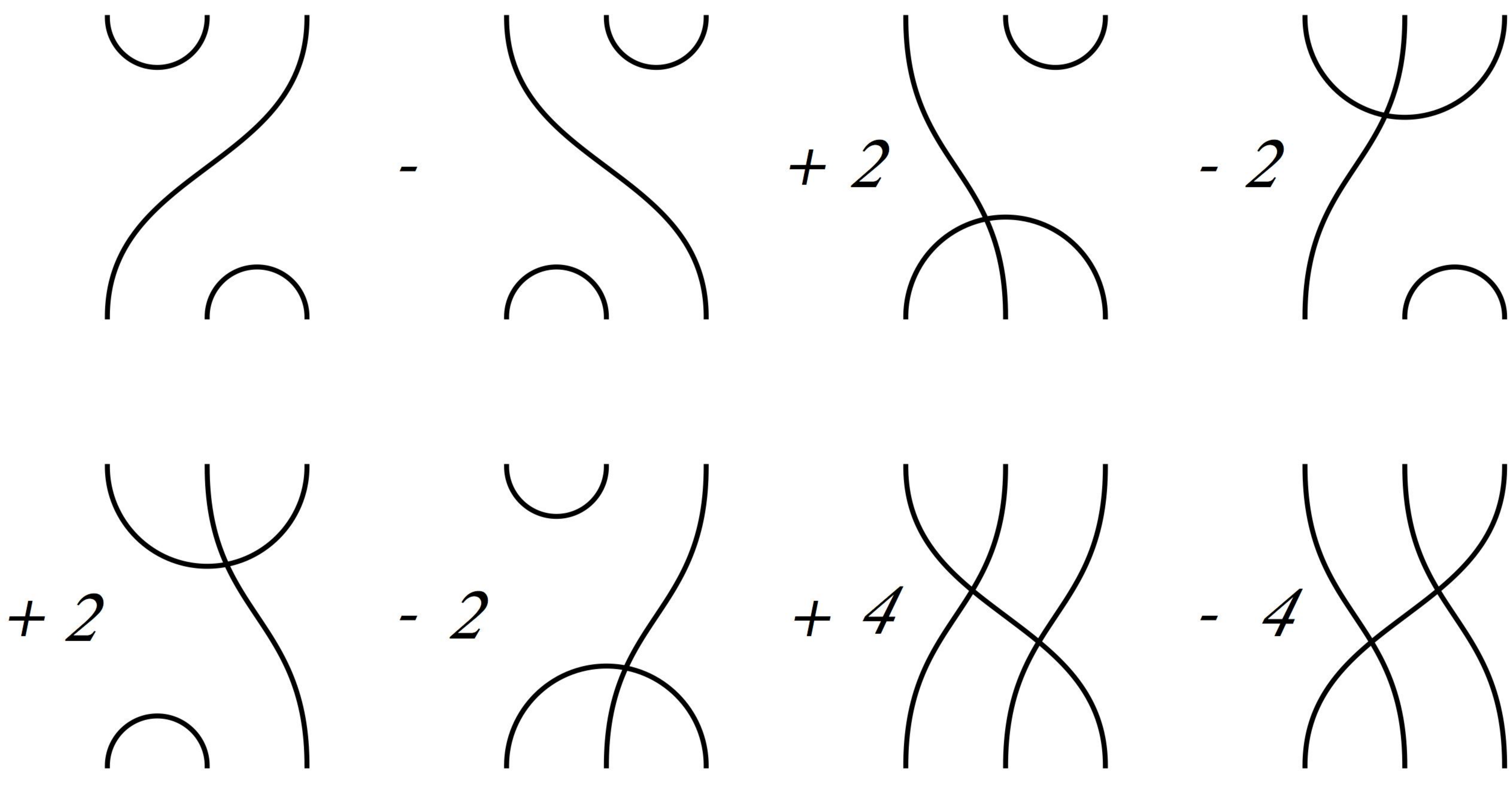}
   \label{fig:subfig2}
 }
\caption{\label{HQ-fig}\small (a) The Hamiltonian (\ref{dilop}). (b) The third charge (\ref{Q3}).}
\end{center}
\end{figure}

\begin{figure}[t]
\begin{center}
 \centerline{\includegraphics[height=4.5cm]{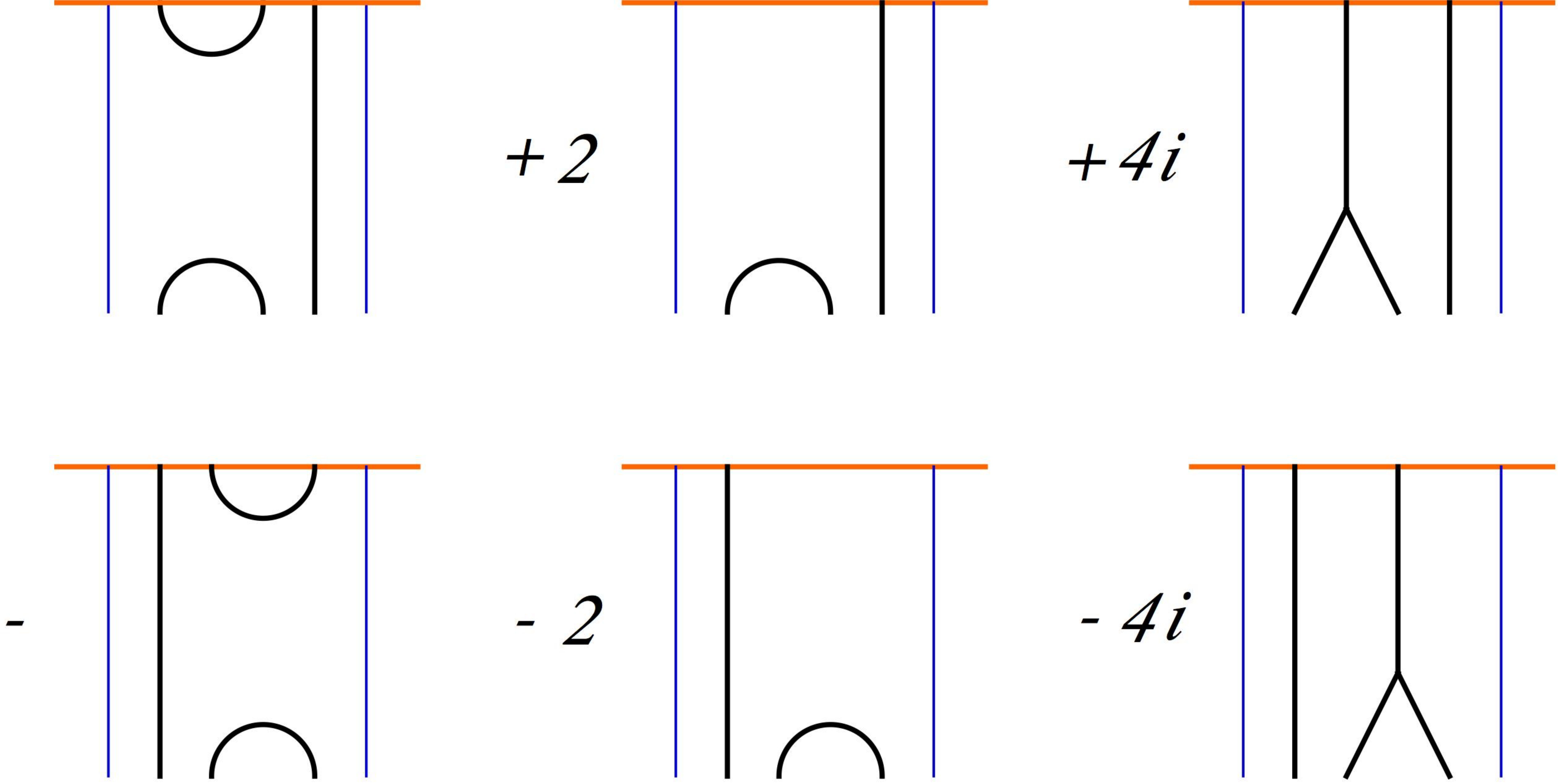}}
\caption{\label{QPsi-fig}\small The result of application of the third charge to the defect state. The horizontal bar denotes the trace over the auxiliary space. The active sites are shown in thick black lines, while the spectator sites, unaffected by $Q$, are shown in blue.}
\end{center}
\end{figure}

\section*{Acknowledgments}

We would like to thank I. Buhl-Mortensen, J.-S.~Caux, A.\ Ipsen, V.~Korepin, I.~Kostov, R.~Nepomechie, J.~Minahan, O.~Ohlsson~Sax, B.~Pozsgay,
G.W.\ Semenoff and P.~Vieira for interesting discussions. K.Z. would like to thank the Galileo Galilei Institute for Theoretical Physics for kind hospitality during the course of this work and the INFN for partial support. C.K.\ and M.d.L.\ were supported in part by FNU through grant number DFF-1323-00082. The work of K.Z. was supported by the Marie
Curie network GATIS of the European Union's FP7 Programme under REA Grant
Agreement No 317089, by the ERC advanced grant No 341222, by the Swedish Research Council (VR) grant
2013-4329, and by RFBR grant 15-01-99504. 

\appendix

\section{Action of third charge on defect state}\label{Q3-appendix}

In this appendix we prove eq.~(\ref{Q3|>=0}). This is most easily done graphically. The Hamiltonian density $H_{lm}$ and the third charge $Q_{lmn}$ are shown in fig.~\ref{HQ-fig}. Applying $Q_{l-1,l,l+1}$ to the defect state $\mathop{\mathrm{tr}}t_{i_1}\ldots t_{i_L}$, we get
\begin{eqnarray}
 (Q\cdot{\rm MPS})_{ijk}&=&\delta _{kj}t_st_st_i-\delta _{ij}t_kt_st_s
+2\delta _{ik}t_jt_st_s-2\delta _{ik}t_st_st_j
\nonumber \\
&&
+2\delta _{ij}t_st_kt_s
-2\delta _{jk}t_st_it_s
+4t_kt_it_j-4t_jt_kt_i,
\nonumber 
\end{eqnarray}
where $i,j,k$ are indices on sites $l-1,l,l+1$ and we have suppressed the rest of the wavefunction unaffected by the operator. Using the commutation relations (\ref{eq:su2relations}) this can be brought to the form
$$
(Q\cdot{\rm MPS})_{ijk}=\delta _{ij}t_st_st_k+2\delta _{ij}t_k+4i\varepsilon _{ijs}t_st_k-\delta _{jk}t_it_st_s-2\delta _{jk}t_i-4it_i\varepsilon _{jks}t_s,
$$
depicted in fig.~\ref{QPsi-fig}. The total charge vanishes upon summation over $l$, which should be clear from the figure.

\bibliographystyle{nb}

\begin{thebibliography}{10}
\ifx\href\asklfhas\newcommand{\href}[2]{#2}\fi
\raggedright
\small
\parskip 0pt

\bibitem{DeWolfe:2001pq}
O.~DeWolfe, D.~Z.~Freedman and H.~Ooguri,
\textit{``{Holography and defect conformal field theories}''},
\textsf{Phys.Rev.~D66,~025009~(2002)},
\href{http://arXiv.org/abs/hep-th/0111135}{\texttt{hep-th/0111135}}.
%
\bibitem{Karch:2000gx}
A.~Karch and L.~Randall,
\textit{``{Open and closed string interpretation of SUSY CFT's on branes with
  boundaries}''},
\textsf{JHEP~0106,~063~(2001)},
\href{http://arXiv.org/abs/hep-th/0105132}{\texttt{hep-th/0105132}}.
%
\bibitem{Nagasaki:2012re}
K.~Nagasaki and S.~Yamaguchi,
\textit{``{Expectation values of chiral primary operators in holographic
  interface CFT}''},
\textsf{Phys.Rev.~D86,~086004~(2012)},
\href{http://arXiv.org/abs/1205.1674}{\texttt{1205.1674}}.
%
\bibitem{Kristjansen:2012tn}
C.~Kristjansen, G.~W.~Semenoff and D.~Young,
\textit{``{Chiral primary one-point functions in the D3-D7 defect conformal
  field theory}''},
\textsf{JHEP~1301,~117~(2013)},
\href{http://arXiv.org/abs/1210.7015}{\texttt{1210.7015}}.
%
\bibitem{Nahm:1979yw}
W.~Nahm,
\textit{``{A Simple Formalism for the BPS Monopole}''},
\textsf{Phys.Lett.~B90,~413~(1980)}.
%
\bibitem{Diaconescu:1996rk}
D.-E.~Diaconescu,
\textit{``{D-branes, monopoles and Nahm equations}''},
\textsf{Nucl.Phys.~B503,~220~(1997)},
\href{http://arXiv.org/abs/hep-th/9608163}{\texttt{hep-th/9608163}}.
%
\bibitem{Giveon:1998sr}
A.~Giveon and D.~Kutasov,
\textit{``{Brane dynamics and gauge theory}''},
\textsf{Rev.Mod.Phys.~71,~983~(1999)},
\href{http://arXiv.org/abs/hep-th/9802067}{\texttt{hep-th/9802067}}.
%
\bibitem{Constable:1999ac}
N.~R.~Constable, R.~C.~Myers and O.~Tafjord,
\textit{``{The Noncommutative bion core}''},
\textsf{Phys.Rev.~D61,~106009~(2000)},
\href{http://arXiv.org/abs/hep-th/9911136}{\texttt{hep-th/9911136}}.
%
\bibitem{Minahan:2002ve}
J.~A.~Minahan and K.~Zarembo,
\textit{``The Bethe-ansatz for {$\mathcal{N}=\mathord{}$4} super Yang-Mills''},
\textsf{JHEP~0303,~013~(2003)},
\href{http://arXiv.org/abs/hep-th/0212208}{\texttt{hep-th/0212208}}.
%
\bibitem{DeWolfe:2004zt}
O.~DeWolfe and N.~Mann,
\textit{``{Integrable open spin chains in defect conformal field theory}''},
\textsf{JHEP~0404,~035~(2004)},
\href{http://arXiv.org/abs/hep-th/0401041}{\texttt{hep-th/0401041}}.
%
\bibitem{klumper1991equivalence}
A.~Kl\"{u}mper, A.~Schadschneider and J.~Zittartz,
\textit{``Equivalence and solution of anisotropic spin-1 models and generalized
  tJ fermion models in one dimension''},
\textsf{J.~Phys.~A~24,~L955~(1991)}.
%
\bibitem{klumper1993matrix}
A.~Kl{\"u}mper, A.~Schadschneider and J.~Zittartz,
\textit{``Matrix product ground states for one-dimensional spin-1 quantum
  antiferromagnets''},
\textsf{Europhys.~Lett.~24,~293~(1993)}.
%
\bibitem{accardi1981topics}
L.~Accardi,
\textit{``Topics in quantum probability''},
\textsf{Phys.~Rept.~77,~169~(1981)}.
%
\bibitem{fannes1989exact}
M.~Fannes, B.~Nachtergaele and R.~Werner,
\textit{``Exact antiferromagnetic ground states of quantum spin chains''},
\textsf{Europhys.~Lett.~10,~633~(1989)}.
%
\bibitem{fannes1992finitely}
M.~Fannes, B.~Nachtergaele and R.~F.~Werner,
\textit{``Finitely correlated states on quantum spin chains''},
\textsf{Comm.~Math.~Phys.~144,~443~(1992)}.
%
\bibitem{ostlund1995thermodynamic}
S.~{\"O}stlund and S.~Rommer,
\textit{``Thermodynamic limit of density matrix renormalization''},
\textsf{Phys.~Rev.~Lett.~75,~3537~(1995)}.
%
\bibitem{rommer1997class}
S.~Rommer and S.~{\"O}stlund,
\textit{``Class of ansatz wave functions for one-dimensional spin systems and
  their relation to the density matrix renormalization group''},
\textsf{Phys.~Rev.~B55,~2164~(1997)}.
%
\bibitem{Alcaraz:2003ya}
F.~C.~Alcaraz and M.~J.~Lazo,
\textit{``{Exact solutions of exactly integrable quantum chains by a matrix
  product ansatz}''},
\textsf{J.Phys.~A37,~4149~(2004)},
\href{http://arXiv.org/abs/cond-mat/0312373}{\texttt{cond-mat/0312373}}.
%
\bibitem{Alcaraz:2006bt}
F.~Alcaraz and M.~Lazo,
\textit{``{Generalization of the matrix product ansatz for integrable
  chains}''},
\textsf{J.Phys.~A39,~11335~(2006)},
\href{http://arXiv.org/abs/cond-mat/0608177}{\texttt{cond-mat/0608177}}.
%
\bibitem{verstraete2008matrix}
F.~Verstraete, V.~Murg and J.~Cirac,
\textit{``Matrix product states, projected entangled pair states, and
  variational renormalization group methods for quantum spin systems''},
\textsf{Adv.~Phys.~57,~143~(2008)}.
%
\bibitem{Katsura:2009vx}
H.~Katsura and I.~Maruyama,
\textit{``{Derivation of Matrix Product Ansatz for the Heisenberg Chain from
  Algebraic Bethe Ansatz}''},
\textsf{J.Phys.~A43,~175003~(2010)},
\href{http://arXiv.org/abs/0911.4215}{\texttt{0911.4215}}.
%
\bibitem{Faddeev:1996iy}
L.~D.~Faddeev,
\textit{``{How Algebraic Bethe Ansatz works for integrable model}''},
\href{http://arXiv.org/abs/hep-th/9605187}{\texttt{hep-th/9605187}}.
%
\bibitem{inverse}
V.~E.~Korepin, N.~M.~Bogolyubov and A.~G.~Izergin,
\textit{``Quantum inverse scattering method and correlation functions''},
Cambridge Univ. Press (1993).
%
\bibitem{Slavnov2007}
N.~A.~Slavnov,
\textit{``The algebraic Bethe ansatz and quantum integrable systems''},
\textsf{Rus.~Math.~Surv.~62,~727~(2007)}.
%
\bibitem{Gaudin:1976sv}
M.~Gaudin,
\textit{``{Diagonalisation d'une Classe d'Hamiltoniens de Spin}''},
\textsf{J.Phys.~France~37,~1087~(1976)}.
%
\bibitem{Korepin:1982gg}
V.~Korepin,
\textit{``{Calculation of norms of Bethe wave functions}''},
\textsf{Commun.Math.Phys.~86,~391~(1982)}.
%
\bibitem{Slavnov:1989}
N.~Slavnov,
\textit{``{Calculation of scalar products of wave functions and form factors in
  the framework of the algebraic Bethe ansatz}''},
\textsf{Theor.Math.Phys.~79,~502~(1989)}.
%
\bibitem{Pozsgay:2009}
B.~Pozsgay,
\textit{``{Overlaps between eigenstates of the XXZ spin-1/2 chain and a class
  of simple product states}''},
\href{http://arXiv.org/abs/1309.4593}{\texttt{1309.4593}}.
%
\bibitem{Brockmann:2014a}
M.~Brockmann, J.~De~Nardis, B.~Wouters and J.-S.~Caux,
\textit{``{A Gaudin-like determinant for overlaps of N\'{e}el and XXZ Bethe
  States}''},
\textsf{J.~Phys.~A:~Math.~Theor.~47,~145003~(2014)},
\href{http://arXiv.org/abs/1401.2877}{\texttt{1401.2877}}.
%
\bibitem{Brockmann:2014b}
M.~Brockmann, J.~De~Nardis, B.~Wouters and J.-S.~Caux,
\textit{``{N\'{e}el-XXZ state overlaps: odd particle numbers and Lieb-Liniger
  scaling limit}''},
\textsf{J.~Phys.~A:~Math.~Theor.~47,~345003~(2014)},
\href{http://arXiv.org/abs/1403.7469}{\texttt{1403.7469}}.
%
\bibitem{Escobedo:2010xs}
J.~Escobedo, N.~Gromov, A.~Sever and P.~Vieira,
\textit{``{Tailoring Three-Point Functions and Integrability}''},
\textsf{JHEP~1109,~028~(2011)},
\href{http://arXiv.org/abs/1012.2475}{\texttt{1012.2475}}.
%
\bibitem{Escobedo:2011xw}
J.~Escobedo, N.~Gromov, A.~Sever and P.~Vieira,
\textit{``{Tailoring Three-Point Functions and Integrability II. Weak/strong
  coupling match}''},
\textsf{JHEP~1109,~029~(2011)},
\href{http://arXiv.org/abs/1104.5501}{\texttt{1104.5501}}.
%
\bibitem{Gromov:2011jh}
N.~Gromov, A.~Sever and P.~Vieira,
\textit{``{Tailoring Three-Point Functions and Integrability III. Classical
  Tunneling}''},
\textsf{JHEP~1207,~044~(2012)},
\href{http://arXiv.org/abs/1111.2349}{\texttt{1111.2349}}.
%
\bibitem{Gromov:2012vu}
N.~Gromov and P.~Vieira,
\textit{``{Quantum Integrability for Three-Point Functions of Maximally
  Supersymmetric Yang-Mills Theory}''},
\textsf{Phys.Rev.Lett.~111,~211601~(2013)},
\href{http://arXiv.org/abs/1202.4103}{\texttt{1202.4103}}.
%
\bibitem{Gromov:2012uv}
N.~Gromov and P.~Vieira,
\textit{``{Tailoring Three-Point Functions and Integrability IV.
  Theta-morphism}''},
\textsf{JHEP~1404,~068~(2014)},
\href{http://arXiv.org/abs/1205.5288}{\texttt{1205.5288}}.
%
\bibitem{Foda:2013nua}
O.~Foda, Y.~Jiang, I.~Kostov and D.~Serban,
\textit{``{A tree-level 3-point function in the su(3)-sector of planar N=4
  SYM}''},
\textsf{JHEP~1310,~138~(2013)},
\href{http://arXiv.org/abs/1302.3539}{\texttt{1302.3539}}.
%
\bibitem{Kazama:2013rya}
Y.~Kazama, S.~Komatsu and T.~Nishimura,
\textit{``{A new integral representation for the scalar products of Bethe
  states for the XXX spin chain}''},
\textsf{JHEP~1309,~013~(2013)},
\href{http://arXiv.org/abs/1304.5011}{\texttt{1304.5011}}.
%
\bibitem{Jiang:2014mja}
Y.~Jiang, I.~Kostov, F.~Loebbert and D.~Serban,
\textit{``{Fixing the Quantum Three-Point Function}''},
\textsf{JHEP~1404,~019~(2014)},
\href{http://arXiv.org/abs/1401.0384}{\texttt{1401.0384}}.
%
\bibitem{Caetano:2014gwa}
J.~Caetano and T.~Fleury,
\textit{``{Three-point functions and $
  \mathfrak{s}\mathfrak{u}\left(1\vert1\right) $ spin chains}''},
\textsf{JHEP~1409,~173~(2014)},
\href{http://arXiv.org/abs/1404.4128}{\texttt{1404.4128}}.
%
\bibitem{Kazama:2014sxa}
Y.~Kazama, S.~Komatsu and T.~Nishimura,
\textit{``{Novel construction and the monodromy relation for three-point
  functions at weak coupling}''},
\textsf{JHEP~1501,~095~(2015)},
\href{http://arXiv.org/abs/1410.8533}{\texttt{1410.8533}}.
%
\bibitem{Jiang:2014cya}
Y.~Jiang, I.~Kostov, A.~Petrovskii and D.~Serban,
\textit{``{String Bits and the Spin Vertex}''},
\textsf{Nucl.Phys.~B897,~374~(2015)},
\href{http://arXiv.org/abs/1410.8860}{\texttt{1410.8860}}.
%
\bibitem{Basso:2015zoa}
B.~Basso, S.~Komatsu and P.~Vieira,
\textit{``{Structure Constants and Integrable Bootstrap in Planar N=4 SYM
  Theory}''},
\href{http://arXiv.org/abs/1505.06745}{\texttt{1505.06745}}.
%
\bibitem{Kazama:2015iua}
Y.~Kazama, S.~Komatsu and T.~Nishimura,
\textit{``{On the singlet projector and the monodromy relation for $psu(2,2|4)$
  spin chains and reduction to subsectors}''},
\href{http://arXiv.org/abs/1506.03203}{\texttt{1506.03203}}.
%
\bibitem{Foda:2011rr}
O.~Foda,
\textit{``{$N=4$ SYM structure constants as determinants}''},
\textsf{JHEP~1203,~096~(2012)},
\href{http://arXiv.org/abs/1111.4663}{\texttt{1111.4663}}.
%
\bibitem{Foda:2012wf}
O.~Foda and M.~Wheeler,
\textit{``{Slavnov determinants, Yang-Mills structure constants, and discrete
  KP}''},
\href{http://arXiv.org/abs/1203.5621}{\texttt{1203.5621}}.
%
\bibitem{Foda:2012yg}
O.~Foda and M.~Wheeler,
\textit{``{Partial domain wall partition functions}''},
\textsf{JHEP~1207,~186~(2012)},
\href{http://arXiv.org/abs/1205.4400}{\texttt{1205.4400}}.
%
\bibitem{Foda:2012wn}
O.~Foda and M.~Wheeler,
\textit{``{Variations on Slavnov's scalar product}''},
\textsf{JHEP~1210,~096~(2012)},
\href{http://arXiv.org/abs/1207.6871}{\texttt{1207.6871}}.
%
\bibitem{Kostov:2012wv}
I.~Kostov and Y.~Matsuo,
\textit{``{Inner products of Bethe states as partial domain wall partition
  functions}''},
\textsf{JHEP~1210,~168~(2012)},
\href{http://arXiv.org/abs/1207.2562}{\texttt{1207.2562}}.
%
\bibitem{Kostov:2012yq}
I.~Kostov,
\textit{``{Three-point function of semiclassical states at weak coupling}''},
\textsf{J.Phys.~A45,~494018~(2012)},
\href{http://arXiv.org/abs/1205.4412}{\texttt{1205.4412}}.
%
\bibitem{Kostov:2012jr}
I.~Kostov,
\textit{``{Classical Limit of the Three-Point Function of N=4 Supersymmetric
  Yang-Mills Theory from Integrability}''},
\textsf{Phys.Rev.Lett.~108,~261604~(2012)},
\href{http://arXiv.org/abs/1203.6180}{\texttt{1203.6180}}.
%
\bibitem{Bettelheim:2014gma}
E.~Bettelheim and I.~Kostov,
\textit{``{Semi-classical analysis of the inner product of Bethe states}''},
\textsf{J.Phys.~A47,~245401~(2014)},
\href{http://arXiv.org/abs/1403.0358}{\texttt{1403.0358}}.
%
\bibitem{Kostov:2014iva}
I.~Kostov,
\textit{``{Semi-classical scalar products in the generalised SU(2) model}''},
\href{http://arXiv.org/abs/1404.0235}{\texttt{1404.0235}}.
%
\bibitem{Bethe:1931hc}
H.~Bethe,
\textit{``{On the theory of metals. 1. Eigenvalues and eigenfunctions for the
  linear atomic chain}''},
\textsf{Z.Phys.~71,~205~(1931)}.
%
\bibitem{Karbach:1998}
{Karbach, M and M\"{u}ller, G},
\textit{``{Introduction to the Bethe ansatz I}''},
\textsf{Comp.~in~Phys.~11,~36~(1998)},
\href{http://arXiv.org/abs/cond-mat/9809162}{\texttt{cond-mat/9809162}}.
%
\bibitem{Beisert:2003tq}
N.~Beisert, C.~Kristjansen and M.~Staudacher,
\textit{``The dilatation operator of {$\mathcal{N}=\mathord{}$4} conformal
  super Yang-Mills theory''},
\textsf{Nucl.~Phys.~B664,~131~(2003)},
\href{http://arXiv.org/abs/hep-th/0303060}{\texttt{hep-th/0303060}}.
%
\bibitem{Sutherland:1995zz}
B.~Sutherland,
\textit{``{Low-Lying Eigenstates of the One-Dimensional Heisenberg Ferromagnet
  for any Magnetization and Momentum}''},
\textsf{Phys.~Rev.~Lett.~74,~816~(1995)}.
%
\bibitem{DharShastry}
A.~Dhar and B.~S.~Shastry,
\textit{``Bloch Walls and Macroscopic String States in Bethe's Solution of the
  Heisenberg Ferromagnetic Linear Chain''},
\textsf{Phys.~Rev.~Lett.~85,~2813~(2000)}.
%
\bibitem{Beisert:2003xu}
N.~Beisert, J.~A.~Minahan, M.~Staudacher and K.~Zarembo,
\textit{``{Stringing spins and spinning strings}''},
\textsf{JHEP~0309,~010~(2003)},
\href{http://arXiv.org/abs/hep-th/0306139}{\texttt{hep-th/0306139}}.
%
\bibitem{Frolov:2003xy}
S.~Frolov and A.~A.~Tseytlin,
\textit{``{Rotating string solutions: AdS/CFT duality in non- supersymmetric
  sectors}''},
\textsf{Phys.~Lett.~B570,~96~(2003)},
\href{http://arXiv.org/abs/hep-th/0306143}{\texttt{hep-th/0306143}}.
%
\bibitem{Kazakov:2004qf}
V.~A.~Kazakov, A.~Marshakov, J.~A.~Minahan and K.~Zarembo,
\textit{``Classical/quantum integrability in AdS/CFT''},
\textsf{JHEP~0405,~024~(2004)},
\href{http://arXiv.org/abs/hep-th/0402207}{\texttt{hep-th/0402207}}.
%
\bibitem{Bissi:2011dc}
A.~Bissi, C.~Kristjansen, D.~Young and K.~Zoubos,
\textit{``{Holographic three-point functions of giant gravitons}''},
\textsf{JHEP~1106,~085~(2011)},
\href{http://arXiv.org/abs/1103.4079}{\texttt{1103.4079}}.
%
\bibitem{Berenstein:1998ij}
D.~E.~Berenstein, R.~Corrado, W.~Fischler and J.~M.~Maldacena,
\textit{``{The operator product expansion for Wilson loops and surfaces in the
  large N limit}''},
\textsf{Phys.~Rev.~D59,~105023~(1999)},
\href{http://arXiv.org/abs/hep-th/9809188}{\texttt{hep-th/9809188}}.
%
\bibitem{Zarembo:2010rr}
K.~Zarembo,
\textit{``{Holographic three-point functions of semiclassical states}''},
\textsf{JHEP~1009,~030~(2010)},
\href{http://arXiv.org/abs/1008.1059}{\texttt{1008.1059}}.
%
\bibitem{Costa:2010rz}
M.~S.~Costa, R.~Monteiro, J.~E.~Santos and D.~Zoakos,
\textit{``{On three-point correlation functions in the gauge/gravity
  duality}''},
\textsf{JHEP~1011,~141~(2010)},
\href{http://arXiv.org/abs/1008.1070}{\texttt{1008.1070}}.
%
\bibitem{Spradlin:2006wk}
M.~Spradlin and A.~Volovich,
\textit{``{Dressing the Giant Magnon}''},
\textsf{JHEP~0610,~012~(2006)},
\href{http://arXiv.org/abs/hep-th/0607009}{\texttt{hep-th/0607009}}.
%
\bibitem{Gomis:2009ir}
J.~Gomis, T.~Okuda and D.~Trancanelli,
\textit{``{Quantum 't Hooft operators and S-duality in N=4 super
  Yang-Mills}''},
\textsf{Adv.Theor.Math.Phys.~13,~1941~(2009)},
\href{http://arXiv.org/abs/0904.4486}{\texttt{0904.4486}}.
%
\bibitem{Gukov:2006jk}
S.~Gukov and E.~Witten,
\textit{``{Gauge Theory, Ramification, And The Geometric Langlands Program}''},
\href{http://arXiv.org/abs/hep-th/0612073}{\texttt{hep-th/0612073}}.
%
\bibitem{Drukker:2008wr}
N.~Drukker, J.~Gomis and S.~Matsuura,
\textit{``{Probing N=4 SYM With Surface Operators}''},
\textsf{JHEP~0810,~048~(2008)},
\href{http://arXiv.org/abs/0805.4199}{\texttt{0805.4199}}.
%
\bibitem{Kapustin:1998fk}
A.~Kapustin and S.~Sethi,
\textit{``{The Higgs branch of impurity theories}''},
\textsf{Adv.Theor.Math.Phys.~2,~571~(1998)},
\href{http://arXiv.org/abs/hep-th/9804027}{\texttt{hep-th/9804027}}.
%
\bibitem{Gaiotto:2008sa}
D.~Gaiotto and E.~Witten,
\textit{``{Supersymmetric Boundary Conditions in N=4 Super Yang-Mills
  Theory}''},
\textsf{J.Statist.Phys.~135,~789~(2009)},
\href{http://arXiv.org/abs/0804.2902}{\texttt{0804.2902}}.
%
\end{thebibliography}

\end{document}